\documentclass[journal]{IEEEtran}
\usepackage{rotating}
\usepackage{amsmath}
\usepackage{bbm}
\usepackage{graphicx}
\usepackage{subcaption}
\usepackage{xcolor}
\usepackage{mathtools}
\usepackage{url}
\DeclareMathOperator*{\argmax}{argmax}

\newcommand{\etabaseline}{\widehat{\eta}_{\rm baseline}}
\newcommand{\etaft}{\widehat{\eta}_{\rm FT}}
\newcommand{\lamtilde}{\widetilde{\lambda}}
\newcommand{\lambar}{\Bar{\lambda}}
\newcommand{\lamhat}{\widehat{\lambda}}
\newcommand{\lammmse}{\widehat{\lambda}_{\rm MMSE}}
\newcommand{\etahat}{\widehat{\eta}}
\newcommand{\varrm}{{\rm var}}
\newcommand{\covrm}{{\rm cov}}
\newcommand{\sigeps}{\sigma_{\epsilon}^2}
\newcommand{\siggams}{\sigma_{\gamma}^2}
\newcommand{\sigxs}{\sigma_x^2}
\newcommand{\etalqm}{\widehat{\eta}_{\rm LQM}}
\newcommand{\etaqm}{\widehat{\eta}_{\rm QM}}
\newcommand{\etatrml}{\widehat{\eta}_{\rm TRML}}
\newcommand{\etaoracle}{\widehat{\eta}_{\rm oracle}}
\newcommand{\etaseq}{\widehat{\eta}_{\rm alt}}
\newcommand{\etaseqhigha}{\widehat{\eta}_{\rm alt,a-high}}
\newcommand{\etaseqlowa}{\widehat{\eta}_{\rm alt,a-low}}
\newcommand{\lamref}{\widehat{\lambda}_{\rm reference}}
\newcommand{\lamseq}{\widehat{\lambda}_{\rm sequential}}
\newcommand{\lamseqONE}{\widehat{\lambda}_{\rm alt}}
\newcommand{\lamseqTWO}{\widehat{\lambda}_{\rm alt,off}}
\newcommand{\lamseqONEhigha}{\widehat{\lambda}_{\rm alt,a-high}}
\newcommand{\lamseqONElowa}{\widehat{\lambda}_{\rm alt,a-low}}

\usepackage{cite}

\begin{document}

\title{Mitigating Current Variation \\ in Particle Beam Microscopy}

\author{Luisa~Watkins, Boston University \\ 
Submitted as a Senior Honors Thesis in Electrical and Computer Engineering \\
June 2021
\thanks{This work was supported in part by the Boston University Undergraduate Research Opportunities Program, a Boston University Clare Boothe Luce Scholar Award, and the US National Science Foundation under Grant No.~1815896.}}

\maketitle

\begin{abstract}
Particle beam microscopy uses a scanning beam of charged particles to create images of samples, and the quality of image reconstruction suffers when this beam current varies over time. Neither conventional reconstruction methods nor time-resolved sensing acknowledges beam current variation, although through sensitivity analysis, my project demonstrates that when the beam current variation is appreciable, time-resolved sensing has significant improvement compared to conventional methods in terms of image reconstruction quality, specifically mean-squared error (MSE). To more actively combat this unknown varying beam current’s effects, my project further focuses on designing an algorithm that uses time-resolved sensing for even better image reconstruction quality in the presence of beam current variation. This algorithm works by simultaneously estimating the unknown beam current variation in addition to the underlying image, offering an alternative to more conventional methods, which exploit statistical assumptions of the image content without explicitly estimating the beam current.
Using a concept of excess MSE due to beam current variation, this algorithm provides a factor of 7 improvement on average, which could lead to less expensive equipment in the future.
Beyond improving the image estimation, this algorithm offers a novel estimation of the beam current, potentially providing more control in manufacturing and fabrication processes. 
\end{abstract}

\IEEEpeerreviewmaketitle

\section{Introduction}
\IEEEPARstart{P}{article} beam microscopy (PBM) is an imaging technique that is used to form images of samples, whether inorganic or biological, at near-atomic resolution. Applied in a wide range of fields, PBM involves a focused beam of charged particles scanning across a sample, pixel by pixel, causing secondary electrons to be emitted in response. These secondary electrons are detected and used to reconstruct an image of the sample, which depicts variation in secondary electron emission per incident particle~\cite{muller1969field}. Within PBM, there are two specific methods I will discuss: \emph{helium ion microscopy} (HIM)~\cite{WardNE:06} and \emph{scanning electron microscopy} (SEM)~\cite{McMullan1995}. While both have uses in imaging biological samples, such as for material science~\cite{phaneuf1999applications}, and nanomachining for manufacturing and fabrication, HIM has the advantage of using heavier helium ions, which allow its reconstructed images to have higher resolution, higher contrast, and higher surface sensitivity~\cite{Notte2016,morgan2006introduction}. On the other hand, SEM uses electrons, which means that its secondary electron yield is lesser. For this reason, I will be analyzing both HIM and SEM with a common abstraction, for values of secondary yield representative of each~\cite{hill2012scanning}. Lastly, although indirect electron detection is more common today, my work assumes direct electron detection since it offers higher signal-to-noise ratio and has been applied to improve imaging resolution in transmission electron microscopy~\cite{peng2020time,mcmullan2009detective,jin2008applications}.

\subsection{Beam Current Variation and Time-Resolved Sensing}
Although the current of incident beam particles is ideally constant, it naturally varies over time. The conventional reconstruction algorithms typically implemented today do not consider that the beam current might vary. These assume that the mean number of particles is constant all throughout the time the sample is being scanned. Under this assumption, instead of using the true beam current, these conventional algorithms use a constant value, which creates stripe artifacts that degrade the performance of the image estimator. 

These conventional algorithms use the standard method of sensing, meaning they detect all the secondary electrons together. Another more recent method of sensing, which acts as an alternative to the standard used, is \emph{time-resolved} (TR) sensing~\cite{PENG2020}. TR sensing works by repeated measurement with low dwell time. While using this type of sensing and its associated group of estimators considerably increases the quality of the image reconstruction, they still assume the incorrect beam current like the conventional algorithms. Although there are fewer artifacts than when conventional methods are used, the TR estimators' reconstructed images still have these stripes present. Without thinking about this natural phenomenon, the image reconstruction algorithm leads to degraded images. The goal of PBM is to take images of samples and learn more about the sample at a near-atomic level. When the images are degraded, it is very difficult to decipher the details of the sample, leaving the PBM process to lose much of its effectiveness.

\subsection{Need for Project}
In today's labs that use microscopy systems and conventional methods, these striped artifacts in the reconstruction can be such a big issue that new images of the samples need to be taken. If the algorithms were better implemented to consider beam current variation, this issue would not arise.  To reduce this, a common solution is to use very precise particle beam generators that have beam current with low variation. Unfortunately, these precise generators are very expensive. While implementing superior algorithms would significantly improve the current setup, the ideal algorithm would not require these expensive generators. In this thesis, I will first discuss the robustness of TR sensing, which demonstrates just how valuable TR sensing is to the world of PBM\@. I then will explain my design of a novel image reconstruction algorithm that leverages TR measurement for even better image reconstruction quality in the presence of beam current variation, allowing for potentially less expensive equipment in the future. My algorithm achieves this by estimating the beam current, which is a novelty itself, possibly contributing to more accurate manufacturing and fabrication.

\section{Measurement Models and Image Estimators}
Now that I have introduced how beam current variation affects the reconstruction of images within PBM, in this section I will discuss the procedure of data collection and processing more in detail, the mathematical models used in these processes, and how various estimators measure the samples using this data. This background material follows the notation of \cite{PENG2020}.

\subsection{Incident Beam Current and Secondary Electron Yield}
As mentioned previously, PBM's main goal is to reconstruct images of samples and it does so by estimating how the sample is affected by incident particles at each pixel of the image. The first step of PBM is to focus a beam of incident particles onto the sample. In reality, the beam generator raster scans the incident beam across the sample, giving a fixed dwell time for each pixel. If we focus on only one pixel, there are a certain number of incident particles hitting the sample. These incident particles are random in nature and follow a Poisson distribution, with mean $\lambda$, which we will call the \emph{dose}. In response to each incident particle $i$, a burst of secondary electrons is emitted. These secondary electrons are also random and follow another Poisson distribution, with mean $\eta$, known as the secondary electron yield. Therefore, for every pixel, there is a value of $\lambda$ and $\eta$.

In order to describe the sample, the goal is to achieve an estimate of $\eta$ for every pixel, creating a complete reconstructed image. If the total number of incident particles at a specific pixel is $M$, then the total number of secondary electrons emitted is simply the sum of the bursts in response to each particle. We will call this total secondary electron count $Y$, which is a compound Poisson random variable and has a \emph{Neyman Type A} distribution~\cite{neyman1939new,teich1981role} with the probability mass function (PMF)
\begin{equation}
\label{eq:neyman}
    P_{Y}(y;\eta,\lambda) = \frac{e^{-\lambda}\eta^y}{y!}\sum^\infty_{m=0}\frac{(\lambda e^{-\eta})m^y}{m!},
\end{equation}
mean
\begin{equation}
\label{eq:mean}
    E[Y] = \lambda\eta,
\end{equation}
and variance
\begin{equation}
\label{eq:variance}
    {\varrm}(Y) = \lambda\eta + \lambda\eta^2.
\end{equation}

\subsection{Formulation of Estimators}
Using the detected secondary electrons $Y$, the sample can be reconstructed by estimating $\eta$ for each pixel in the image. The most straightforward approach is the \emph{conventional} estimator, which uses the relation in \eqref{eq:mean} to its advantage. Therefore, the baseline \emph{conventional} estimator is
\begin{equation}
\label{eq:conv}
    \etabaseline(Y) = \frac{Y}{\lambda},
\end{equation}
with \emph{mean-squared error}
\begin{equation}
\label{eq:conv_mse}
     {\rm MSE}\big(\etabaseline\big)=\frac{\eta(1+\eta)}{\lambda}.
\end{equation}

Another method of sensing is TR sensing, which uses repeated measurement with low dwell time. Instead of using the one value of $Y$, TR sensing involves using a vector of secondary electron values for each pixel. If we focus on one pixel once again, TR sensing splits the dwell time into $n$ sub-acquisitions, giving $n$ measurements: $Y_1, Y_2, ... Y_n$. If $n$ is large enough, then the dose per sub-acquisition $\lambda/n$ is so small that it is unlikely that more than one incident particle will be observed~\cite{PENG2020}. This means that the estimators using this method of sensing can have an idea of the number of incident particles $M$. This gives a much better estimate than simply having an idea of the dose $\lambda$, especially if only the mean of $\lambda$, $\lamtilde$, is known.

There are multiple estimators associated with TR sensing. The first is the \emph{quotient mode} (QM) estimator
\begin{equation}
\label{eq:qm}
    \etaqm = \frac{Y_1+Y_2+\cdots+Y_n}{\sum^n_{k=1}\mathbbm{1}_{\{Y_k>0\}}},
\end{equation}
which uses the indicator function $\mathbbm{1}_{\{Y_k>0\}}$ to count how many of the $n$ sub-acquisitions have secondary electrons in them. By dividing $Y$ by this indicator count, this estimator is analogous to the baseline estimator \eqref{eq:conv}, but instead of using $\lamtilde$, the QM estimator \eqref{eq:qm} uses an estimate of $M$. Another estimator associated with TR sensing is similar to the QM estimator, but considers that when the secondary electron yield is low, a large bias appears in the QM estimator. This is the \emph{Lambert quotient mode} (LQM) estimator, with closed form solution
\begin{equation}
    \label{eq:lqm}
    \etalqm = W(-\etaqm e^{-\etaqm}) + \etaqm,
\end{equation}
where $W(\cdot)$ is the Lambert W function~\cite{corless1996lambertw}. The last estimator associated with TR sensing that I will discuss uses the joint PMF of the vector of detected secondary electrons. Therefore, the  \emph{time-resolved maximum likelihood} (TRML) estimator is
\begin{equation}
    \label{eq:trml}
    \etatrml = \argmax_{\eta}\prod^n_{k=1}P_{Y}(y_k;\eta,\lambda/n),
\end{equation}
which can be performed using a grid search. While the QM and LQM estimators are independent of $\lambda$, the TRML estimator proves to provide the most accurate image in reconstruction quality~\cite{peng2020time,PENG2020}. For all of my tests, I will be using the \emph{mean-squared error} (MSE), as it is the accepted error metric for image accuracy.

\begin{figure*}
    \centering
    \begin{subfigure}{0.32\linewidth}
      \includegraphics[width=\linewidth]{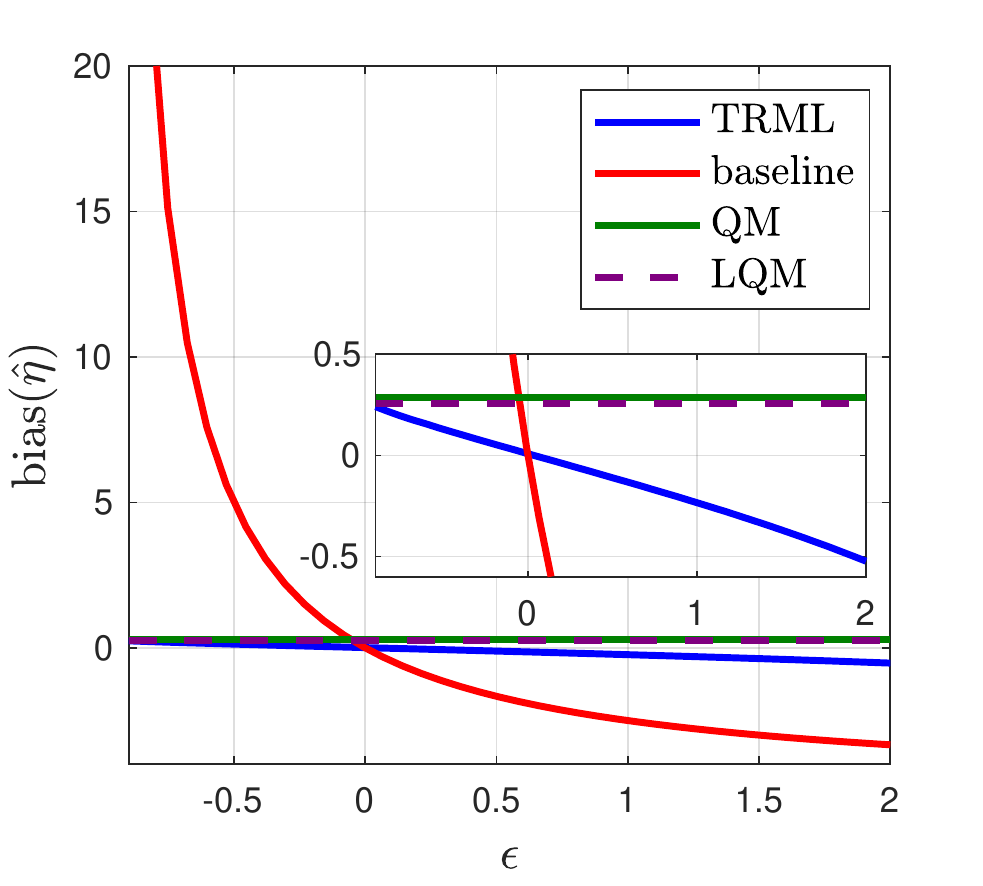}
      \captionsetup{justification=centering}
      \caption{\small Bias across $\epsilon$}
      \label{fig:bias}
    \end{subfigure}
    \begin{subfigure}{0.32\linewidth}
      \includegraphics[width=\linewidth]{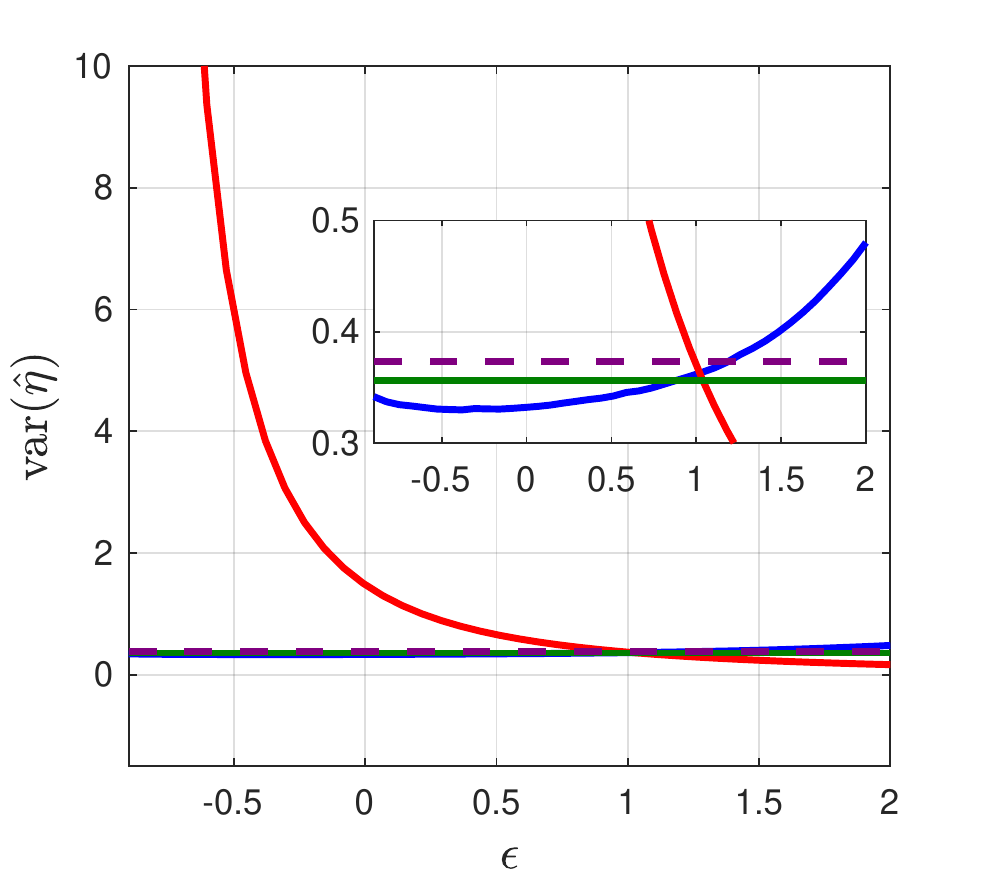}
      \captionsetup{justification=centering}
      \caption{\small Variance across $\epsilon$ }
      \label{fig:var}
    \end{subfigure}
    \begin{subfigure}{0.32\linewidth}
      \includegraphics[width=\linewidth]{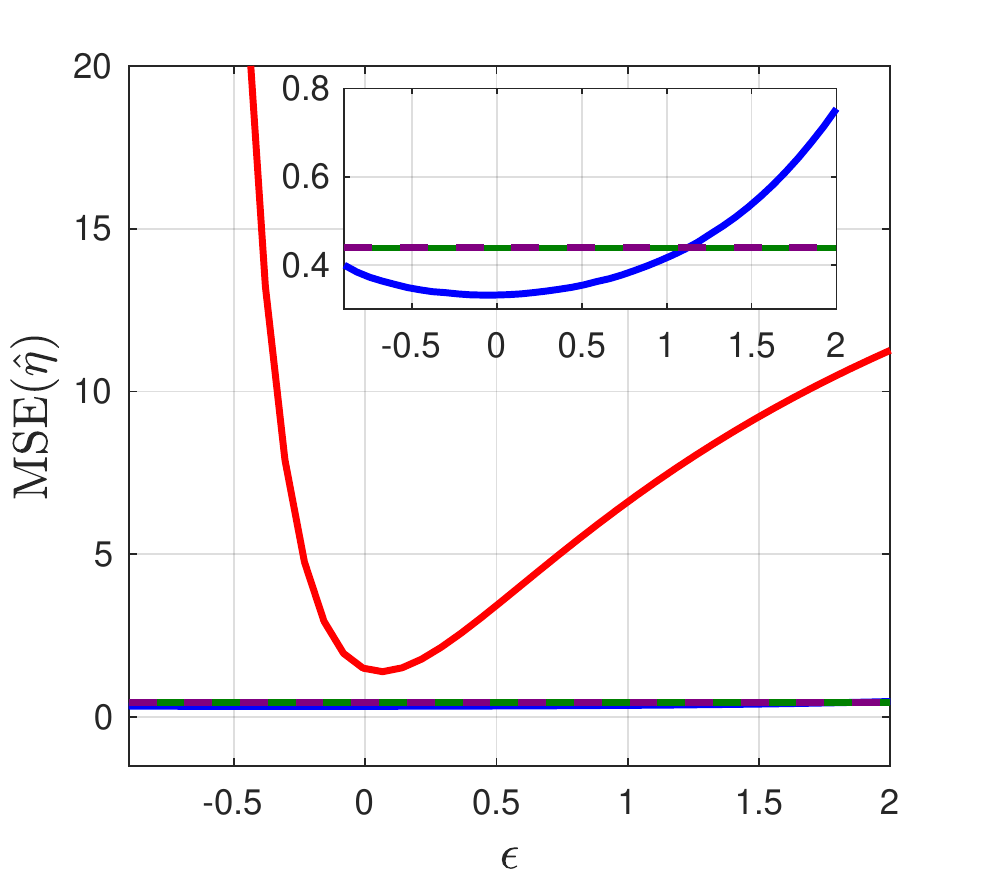}
      \captionsetup{justification=centering}
      \caption{\small  MSE across $\epsilon$ }
      \label{fig:mse}
    \end{subfigure}
    \caption{ Bias, variance, and MSE of $\hat{\eta}$ as a function of $\epsilon$ when $\eta = 5$, $\lambda = 20$, and $n = 200$.}
    \label{fig:biasVarMSE}
\end{figure*}

\begin{figure*}
    \centering
    \begin{subfigure}{0.39\linewidth}
      \includegraphics[width=\linewidth]{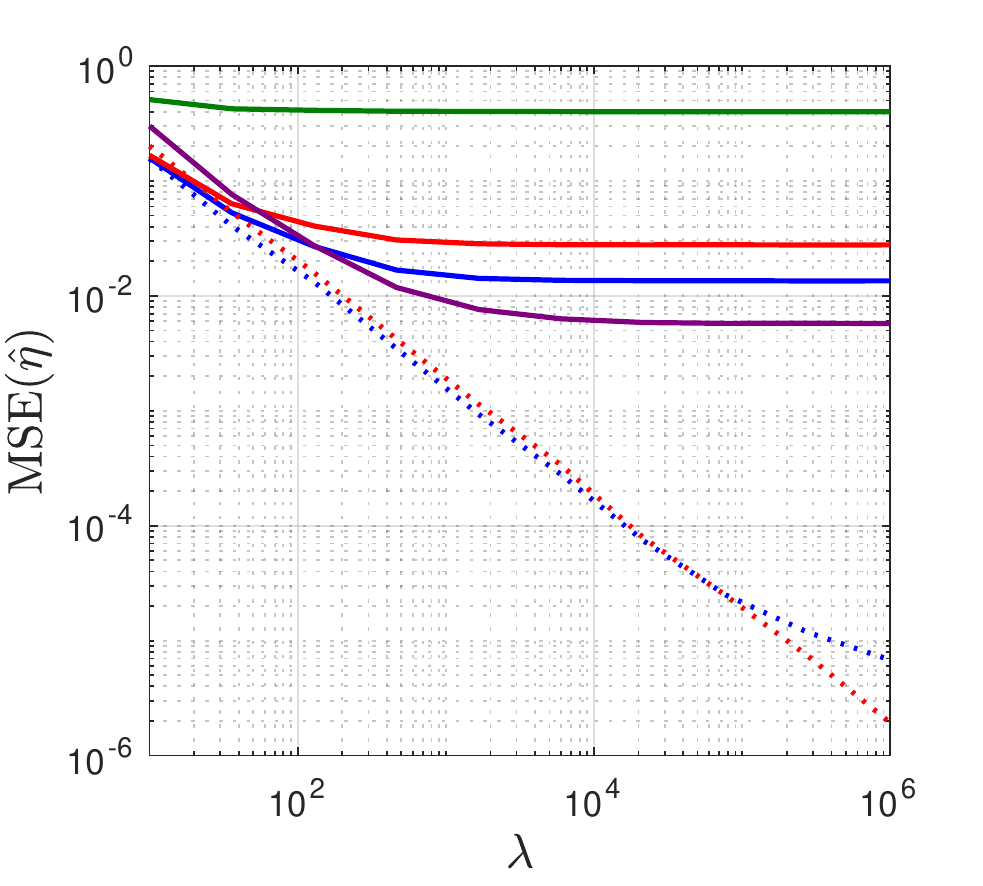}
      \captionsetup{justification=centering}
      \caption{$\eta = 1$}
      \label{fig:loweta}
    \end{subfigure} 
    \begin{subfigure}{0.39\linewidth}
      \includegraphics[width=\linewidth]{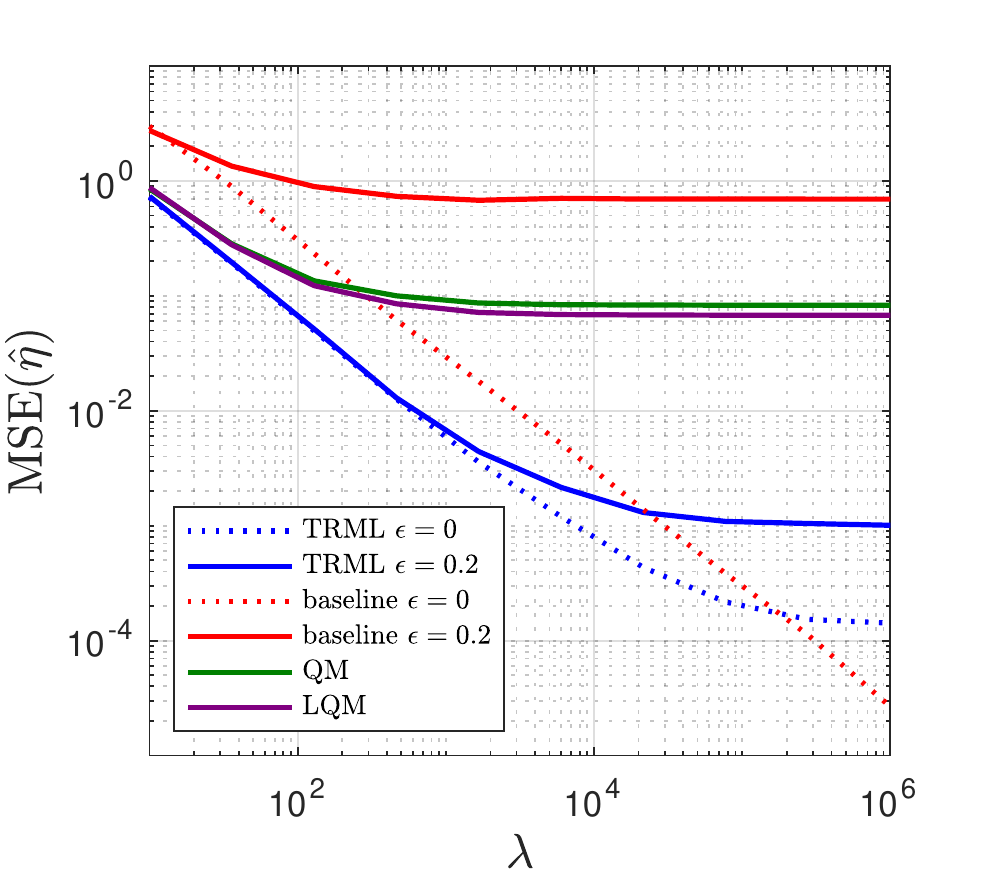}
      \captionsetup{justification=centering}
      \caption{$\eta = 5$}
      \label{fig:higheta}
    \end{subfigure} 
    \caption{MSE of $\etahat$ as a function of total dose $\lambda$ when $\lambda/n = 0.1$. Note the log-log scale.
    }
    \label{fig:mse_v_lam}
\end{figure*}

\section{Robustness of Time-Resolved Measurement}
\label{section:robust}
In the previous section, I did not not mention the beam current variation, simply because none of these estimators, even those associated with TR sensing, consider it. This means that instead of using the true dose $\lambda$, the baseline and TRML estimators use an incorrect value, which is assumed to be accurate.

If we focus on one pixel once more, we can think of our true dose as $\lambda$ and our estimator dose as $\lambar = \lambda(1+\epsilon)$, which represents that the estimator's knowledge of the dose is a bit off of the true $\lambda$ value. The baseline estimator has an MSE of
\begin{equation}
\label{eq:mse}
    {\rm MSE}\big(\etabaseline(\lambar)\big)=\eta^2\Big(\frac{\epsilon}{1+\epsilon}\Big)^2 + \frac{\eta(1+\eta)}{\lambda(1+\epsilon)^2}.
\end{equation}
By scanning across a range of $\epsilon$ values, we can see how each of the estimators perform as their accuracy of $\lambda$ changes. In Figure~\ref{fig:biasVarMSE}, the MSE, bias, and variance for all the estimators is plotted across this range, for the case of $\eta = 5$, $\lambda = 20$, and $n = 200$. As seen in the plot, the baseline estimator has the worst performance across the entire range, proving how ineffective this estimator can be, even when using the correct dose value. By looking at Figure~\ref{fig:bias}, we can see that as $\epsilon$ tends away from 0 in either direction, the absolute value of the bias drastically increases, making it an important factor leading to the poor performance of the baseline estimator when beam current is imperfectly known.

The estimators associated with TR sensing do a tremendous job across this range, especially the QM and LQM estimators, which have a constant MSE due to their independence of $\lambda$. It is important to note that the QM and LQM manage to perform better than the baseline, even whether they have \emph{no} knowledge of $\lambda$ and the baseline has \emph{perfect} knowledge of it.

In terms of the TRML estimator, even though it does depend on $\lambda$, it does so much less than the baseline, allowing it to have the best performance for most values of $\epsilon$. The TRML is only beaten by the QM and LQM for large magnitudes of $\epsilon$. Even so, we can see by examining Figure~\ref{fig:mse} that all of the estimators associated with TR measurement are extremely robust to beam current variation. The robustness of these TR-associated estimators is further explored in Section~\ref{section:simulate}, when they are tested not just at one pixel, but across an entire sample.

In terms of the baseline estimator, the MSE in (\ref{eq:mse}) and that in (\ref{eq:conv_mse}) differ only in what dose value is used in the estimator. Since (\ref{eq:mse}) uses the wrong dose value to estimate, there is an extra term in its MSE expression, which demonstrates that as $\eta$ increases, the MSE of this estimator has a large growing term that (\ref{eq:conv_mse}) does not have; this is the \emph{bias}. Besides this extra term, both (\ref{eq:conv_mse}) and (\ref{eq:mse}) have the same inverse relationship with $\lambda$; as $\lambda$ grows, the MSE lowers. As seen in Figure~\ref{fig:mse_v_lam}, the baseline estimator is able to outperform the other TR-associated estimators, but only when the correct dose is used to estimate and when $\lambda$ is large enough. If the wrong dose value is used, then the bias dominates, leading to a very large MSE\@. In addition, since this bias term is heavily dependent on $\eta$, we see that by using a larger value of $\eta$, the baseline estimator struggles more to outperform the other estimators. This can be seen by comparing Figures~\ref{fig:loweta} and~\ref{fig:higheta}.

\section{Dose Estimation}
\label{section:design}
While using estimators that are robust to beam current variation does significantly improve image quality, they still use the wrong dose value. It is quite remarkable that the TR-associated estimators do so well while using the incorrect dose, but if they did have a better estimate of $\lambda$, the image quality could improve even more, pushing the limits of what is achievable. In this section, I will explain a new algorithm that not only estimates $\eta$, but also $\lambda$, by alternating between estimating these two; by using a more accurate value of $\lambda$, a more accurate value of $\eta$ can be estimated. In addition, estimating $\lambda$ itself has its own benefits. As mentioned in the introduction, PBM is useful not just for imaging, but also for nanomachining purposes. Therefore, by having more knowledge of the incident dose than simply a constant value, these processes can be more precisely executed. 

\subsection{Model for Beam Current}
\label{section:modeldescription}
In order to design an algorithm that takes the beam current variation into account, a model is used. Since this beam current varies over time, I will use an \emph{autoregressive} (AR) model~\cite{vetterli2014foundations}, specifically an AR(1) process,
\begin{equation}
    \label{eq:model}
    \lambda_n = x_n + a\lambda_{n-1} + c,
\end{equation}
where $a$ and $c$ are parameters that can be altered to simulate different conditions, and $x_n \sim N(0,\sigxs)$, meaning Gaussian white noise with zero mean and variance $\sigxs$. 
Throughout the rest of the paper, $\lamtilde$ will be used to signify the value of $\lambda$ that is assumed by the estimator, instead of using the true dose value of $\lambda$ at each pixel.
Additionally, the AR model varies around this nominal value of $\lamtilde$, therefore making it the mean of $\lambda$; thus, $c = \lamtilde(1-a)$.
A derivation of this is found in the Appendix. 
The parameter $a$ is the correlation coefficient and signifies how similar nearby values of $\lambda$ are to each other. As $a$ increases, the values are more correlated, meaning $\lambda$ varies slower over the pixels, making the striping artifacts thicker vertically. As $a$ decreases, there is less correlation between $\lambda$ values, meaning it varies faster over the pixels, making the striping not extend to the width of the image, but different striping sections within even one row of pixels in the image. Lastly, $\sigxs$ contributes to how much the values of $\lambda$ vary from their mean $\lamtilde$.

\subsection{Sequential Filter Estimator for $\lambda$}
The biggest contribution my algorithm offers is a way to estimate $\lambda$. In this subsection, I will explain the piece of my algorithm that does this estimation, and then in the following subsection, I will review the algorithm in its entirety.

When constructing this $\lambda$ estimator, the idea was to ideally use an estimator with the lowest MSE, otherwise known as the \emph{minimum mean-squared error} (MMSE) estimator, or $\lammmse$~\cite{yates2014probability}. Unfortunately, using $\lammmse$ proved to be much too complicated to compute. To see why, we can inspect the case where $\eta$ is known. In this scenario, we know that
\begin{equation}
\label{eq:mmse}
    \lammmse = E[\lambda|Y=y],
\end{equation}
where $f_{\lambda|Y}(\lambda|y)$ is used:
\begin{equation}
\label{eq:pdf_lambdagiveny}
    f_{\lambda|Y}(\lambda|y) = \frac{P_{Y|\lambda}(y|\lambda)f_{\lambda}(\lambda)}{P_Y(y)}.
\end{equation}
Although both $P_{Y|\lambda}(y|\lambda)$ and $f_{\lambda}(\lambda)$ are known from (\ref{eq:neyman}) and $\lambda \sim N(\lamtilde,\sigma_{\lambda}^2)$, respectively, (\ref{eq:pdf_lambdagiveny}) would require an integral, in addition to the integral needed for the expected value computation of (\ref{eq:mmse}). 

Therefore, while ideal, using the MMSE estimate is too complicated, even for this scenario where $\eta$ is known, let alone when unknown; for this reason,  a linear estimator will be used. This linear $\lambda$ estimator is the \emph{linear mean-squared error} (LMSE) estimator~\cite{yates2014probability}. Similarly to (\ref{eq:mmse}), this estimator is a function of $Y$; this decision makes the most sense since this estimator uses what is directly proportional to $\lambda$, helping to give the best estimate of $\lambda$.

In terms of this LMSE $\lambda$ estimator, it moves sequentially through the image's pixels, indexed by p. In the overall algorithm, the sequential filter estimator is applied both forward and backward.
Although linearly dependent on $Y$, this designed estimator is also allowed to depend on $\eta$ and the previous pixel's estimate of $\lambda$. Using the optimum LMSE estimator~\cite{yates2014probability} of $\lambda$ given $Y$, the sequential filter is
\begin{equation}
    \label{eq:filter}
    \lamhat_p(y_p) = \frac{{\rm cov}(\lambda_p,y_p)}{{\rm var}(y_p)}(y_p - E[y_p]) + E[\lambda_p],
\end{equation}
where the expectations, variance, and covariance terms are dependent on $\eta_p$ and $\lambda_{p-1}$. If the true $\eta_p$ and $\lambda_{p-1}$ are somehow known, then the sequential filter would be (\ref{eq:filter}), where
\begin{align}
    \label{eq:cov-ideal}
    {\rm cov}(\lambda_p,y_p) &= \eta_p\sigxs, \\
    \label{eq:var_y-ideal}
    {\rm var}(y_p) &= (a\lambda_{p-1} + c)(\eta_p + \eta_p^2) + \eta_p^2\sigxs, \\
    \label{eq:mean_y-ideal}
    E[y_p] &= \eta_p(a\lambda_{p-1}+c), \\
    \label{eq:mean_lambda-ideal}
    E[\lambda_p] &= a\lambda_{p-1} + c.
\end{align}

In reality, these terms within (\ref{eq:filter}) use estimates and not the true values of $\eta_p$ and $\lambda_{p-1}$. By assuming these estimates are not perfect, (\ref{eq:filter}) can be a more accurate estimator. For this design, I made the assumption that the errors of $\lambda_{p-1}$ and $\eta_p$ can be treated as zero-mean Gaussian. For $\lambda$, instead of assuming the ideal  $\lamhat_{p-1} = \lambda_{p-1}$, we have $\lamhat_{p-1} = \lambda_{p-1} + \epsilon_p$, where $\epsilon_p~\sim~N(0,\sigeps)$. Similarly, there is no possible way to acquire the true $\eta$, which means that there must be some noise on $\etahat$. Instead of the ideal $\etahat_p = \eta_p$, we have $\etahat_p = \eta_p + \gamma_p$, where $\gamma_p \sim N(0,\siggams)$. Although not ideal, there will always be error when estimating $\eta$, meaning that $\siggams > 0$.

Therefore, when the proper assumptions are made, the sequential filter is (\ref{eq:filter}), where 
\begin{align}
    \label{eq:cov}
    {\rm cov}(\lambda_p,y_p) &= \etahat_p(\sigxs + a^2\sigeps), \\
    \label{eq:var_y}
    {\rm var}(y_p) &= (a\lamhat_{p-1} + c)(\etahat_p + \etahat_p^2 + \siggams) \nonumber \\
    & \quad + \siggams(\sigxs + a^2\sigeps + a^2\lamhat_{p-1}^2 + 2ac\lamhat_{p-1} + c^2) \nonumber \\
    & \quad + \etahat_p^2(\sigxs + a^2\sigeps), \\
    \label{eq:mean_y}
    E[y_p] &= \etahat_p(a\lamhat_{p-1}+c), \\
    \label{eq:mean_lambda}
    E[\lambda_p] &= a\lamhat_{p-1} + c.
\end{align}
The full derivation of this sequential filter can be found in the Appendix, resulting in (\ref{eq:finalfilter}), its full form. Importantly, any $\eta$ estimator can be used as input for the $\lambda$ estimation, as long as an appropriate $\siggams$ is chosen. Regardless of how $\eta$ is estimated, this sequential filter uses that existing $\etahat$ to produce a $\lamhat$.

The last piece to consider when designing this estimator is how to select the parameters, $\sigeps$ and $\siggams$. In other words, we need a way to describe how much the estimates of $\lambda$ and $\eta$ vary from their true values. These parameters are selected at the beginning of each iteration of $\lambda$ estimation within the full algorithm.

\subsubsection{Selecting $\sigeps$}
By assuming that both $\sigeps = {\rm MSE}(\lamhat_p)$ and the $\lambda$ estimator is unbiased, and $\sigeps$ is constant across the entire image, we can solve for the parameter needed. We have
\begin{equation}
    \label{eq:mse_lamhat}
    {\rm MSE}(\lamhat_p) = \varrm(\lambda_p) - \frac{\covrm(\lambda_p,y_p)^2}{\varrm(y_p)},
\end{equation}
where
\begin{equation}
    \label{eq:var_lambda}
    \varrm(\lambda_p) = \sigxs + a^2\sigeps,
\end{equation}
and $\covrm(\lambda_p,y_p)$ is \eqref{eq:cov} and $\varrm(y_p)$ is \eqref{eq:var_y}. By taking the average value of $\etahat$ and $\lamhat$ across all the pixels, we can solve for one value of $\sigeps$ for the entire image.

To be more clear, the same value of $\sigeps$ is used for each pixel in the sequential filter. Although each estimate of $\lambda$ is technically different, including its error from the true $\lambda$, computing one value of $\sigeps$ for every pixel proved to be too computationally expensive for not that much benefit. Through testing, it was found that there was not much of a difference whether one constant value of $\sigeps$ was used or a different value for each pixel was used. Therefore, since computing one value is much less expensive, this was what was ultimately chosen for the design.

\subsubsection{Selecting $\siggams$}
As mentioned previously, how $\etahat$ is acquired is completely independent of how the sequential filter works. No matter which $\etahat$ is used, the sequential filter can run the same way. Although not a requirement, the complete algorithm, which will be explained in the next section, uses $\etatrml$ as its estimate of $\eta$.

In Figure 5b of \cite{peng2020time}, we can see that the bias of $\etatrml$ is very small, even when comparing against other TR-associated estimators. Using this fact to assume that $\etatrml$ is unbiased, this means that $\siggams = \rm MSE(\etatrml)$. This is due to the fact that $\rm bias = 0$, making $\rm MSE = variance$, which is the variance of $\gamma$, or $\siggams$. Similarly to $\sigeps$, one constant value of $\siggams$ is found to keep things simple. To get this parameter, a data set of estimated $\rm MSE(\etatrml)$ values, dependent on $\lambda$, is used, courtesy of Minxu Peng. To use this data set, we use $\lamtilde$ and the average value of $\etatrml$ across the whole image so we can scan through the data to find the closest matching pairing, giving us a good value of $\siggams$ to use for all the pixels.

\begin{figure*}
    \centering
    \begin{subfigure}{0.39\linewidth}
      \includegraphics[width=\linewidth]{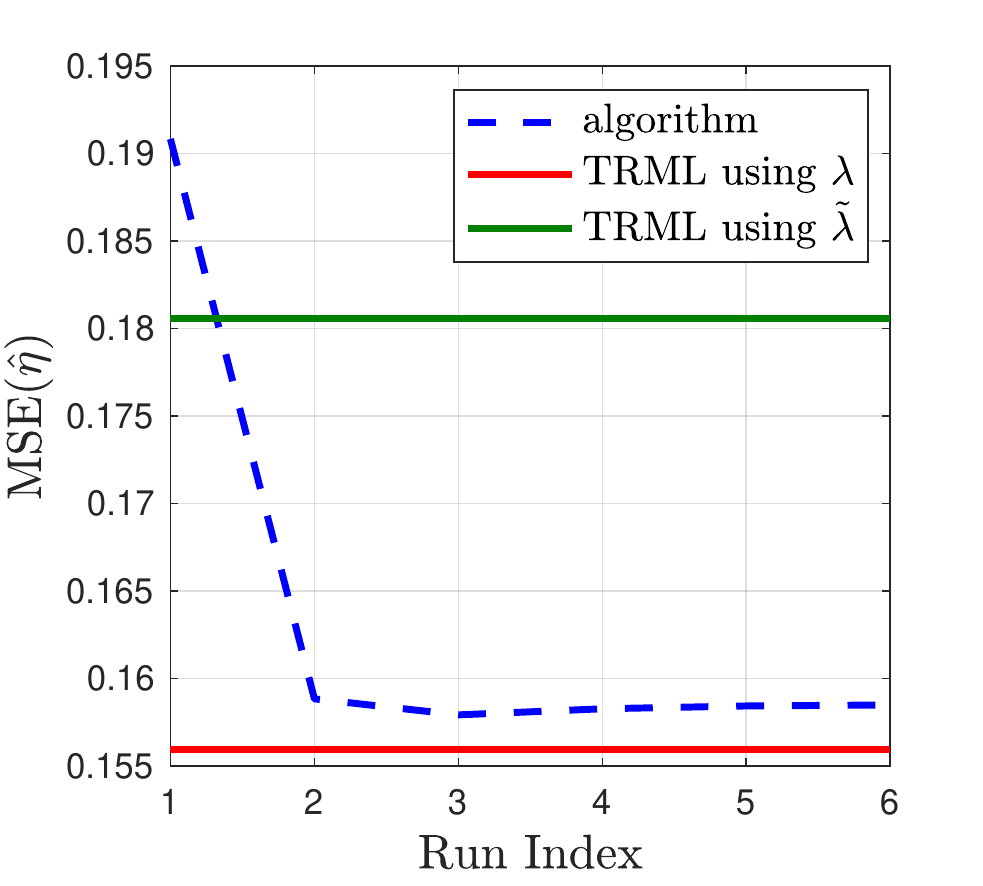}
      \captionsetup{justification=centering}
      \caption{\small MSE($\etahat$) across iterations}
      \label{fig:iterations}
    \end{subfigure} 
    \begin{subfigure}{0.39\linewidth}
      \includegraphics[width=\linewidth]{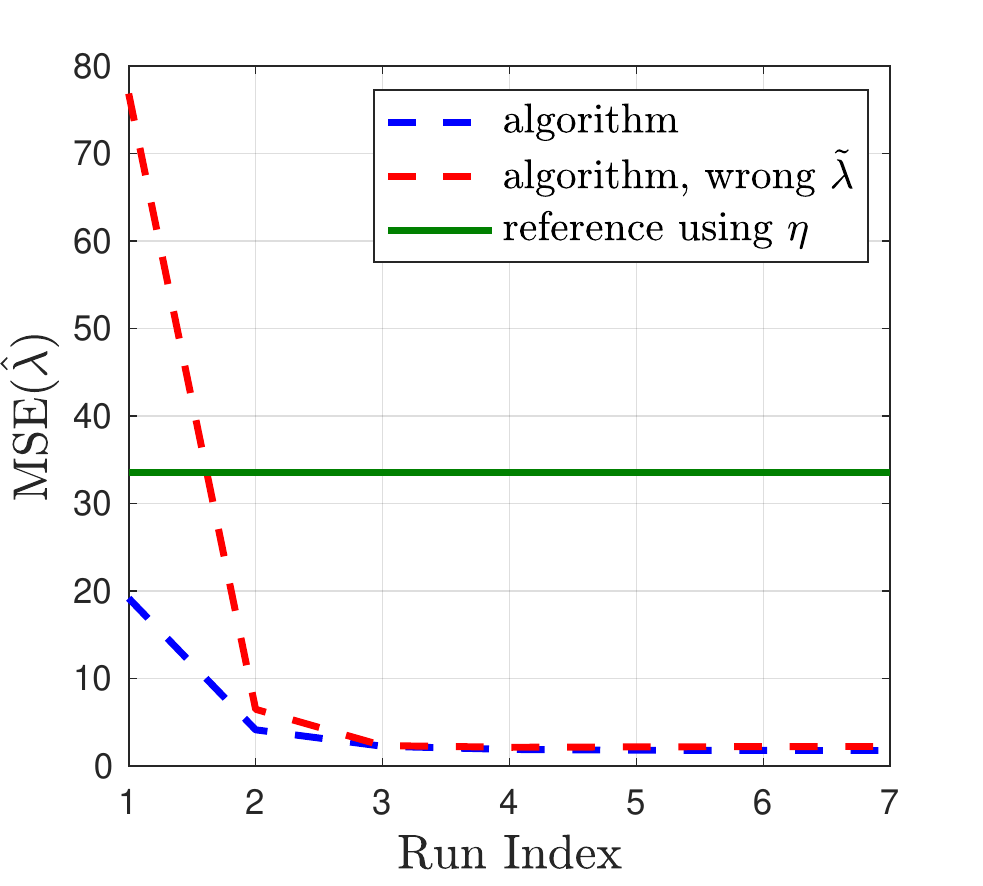}
      \captionsetup{justification=centering}
      \caption{\small MSE($\lamhat$) across iterations}
      \label{fig:literations}
    \end{subfigure} 
    \caption{MSE of $\etahat$ and $\lamhat$ across iterations of the new proposed algorithm, in comparison to TRML and reference estimators, respectively. These plots correspond to the data of Example 1 in Figure~\ref{fig:micrographs1}.}
    \label{fig:mse_v_index}
\end{figure*}

\subsection{Complete Algorithm}
\label{section:algorithm}
With the sequential filtering $\lamseq$ (\ref{eq:filter}) explained, we can move forward to the alternating algorithm.
The main idea of this algorithm is to alternate between estimating $\eta$ and $\lambda$, completing one and then using its estimate to complete the other and so on. The $\eta$ estimate comes from the TRML estimator, whose inputs are $\lamhat$, the time-resolved vector of secondary electrons $[Y_1,Y_2,...Y_n]$ for every pixel, the number of sub-acquisitions $n$, and whose output is obviously $\etatrml$. The $\lambda$ estimate comes from the sequential filter, whose inputs are $\lamhat$ and the parameters from the AR model, $\etatrml$, $Y$ for every pixel, and the number of pixels, and whose output is $\lamseq$.

From looking at the inputs for each estimator, we can see that an estimate of one parameter is dependant on the estimate of the other; i.e. to estimate $\eta$, we need an estimate of $\lambda$, and vice versa. To get the best possible estimate of $\eta$, we can alternate between $\etatrml$ \eqref{eq:trml} and $\lamseq$ \eqref{eq:filter} one at a time. First, $\eta$ is estimated using TRML, where $\lamhat$ is simply $\lamtilde$, which we know to be the mean of $\lambda$. Then, $\lambda$ is estimated using the sequential filter, where $\etahat$ is $\etatrml$. This leads to the next iteration of $\etatrml$ and $\lamseq$ using $\lamseq$ as input instead of $\lamtilde$ as $\lamhat$.

This pattern of back and forth continues until convergence, which can be seen in Figure~\ref{fig:mse_v_index}. In these plots, we can see how the algorithm performs over time in both estimation of $\eta$ and $\lambda$. In Figure~\ref{fig:iterations}, we inspect the performance of the algorithm's $\eta$ estimate, $\etaseq$, over 6 iterations, compared to $\etatrml$ using $\lambda$ and $\etatrml$ using $\lamtilde$. As seen in this plot, the algorithm does a tremendous job at improving its performance, closely approaching what would be capable if \emph{perfect} knowledge of the dose $\lambda$ were known. In Figure~\ref{fig:literations}, we inspect the performance of the algorithm's $\lambda$ estimate, $\lamseqONE$, and $\lamseqTWO$ over the 7 iterations, compared to the reference $\lambda$ estimator
\begin{equation}
    \label{eq:lam_ref}
    \lamref = \frac{Y}{\eta}.
\end{equation}
In this case, $\lamseqONE$ is the $\lambda$ estimate that corresponds with $\etaseq$, which starts with $\lamtilde$ as its estimate, whereas $\lamseqTWO$ is another test of the algorithm, which starts with an incorrect $\lamtilde$ value that is 2 standard deviations below the true mean value. This means that there is a mismatch between the $\lamtilde$ assumed by the estimator and the $\lamtilde$ used as the mean of the AR model. 
In these tests, $\lamref$ does use the true $\eta$, in order to give it its best chance against other $\lambda$ estimators that have some knowledge of the dose. This is why $\lamseqTWO$ is also tested, so that we can demonstrate that even when the algorithm has an extremely wrong starting estimate of the dose, it is still able to achieve great performance. This is indeed what Figure~\ref{fig:literations} shows, where although $\lamseqONE$ and $\lamseqTWO$ are initialized with very different values of dose, they are both able to achieve around the same MSE\@.

In terms of how long to run this algorithm, or how many iterations are needed to reach convergence, this depends on the difficulty of the problem. As seen in Figure~\ref{fig:mse_v_index}, this specific case needed about 6 iterations, or even fewer. This algorithm needs to run longer for when $\eta$ values are smaller; although the MSE may be lower, smaller values of $\eta$ make it harder to improve upon $\etatrml$ due to the the fewer secondary electrons detected, meaning there is less information to work with.

\section{Simulation Comparisons}
\label{section:simulate}
Now that this new design algorithm has been explained, we can go into its performance for multiple different testing scenarios. In addition to this algorithm, other methods will be tested, in order to have adequate comparisons.

\begin{figure*}
    \centering
    \begin{minipage}{1\textwidth}
        \text{\normalsize \textbf{Example 1}: HIM-\emph{medium} with $\eta \in [1,\,5]$, $\tilde{\lambda} = 20$, $\sigma_{\lambda}/\tilde{\lambda} = 0.2$, and $a = 0.999$.}
    \end{minipage}
    \begin{minipage}{     .17\linewidth}
        \centering
        \vspace{1mm}
        \text{\textbf{\normalsize $\eta$}}
        \includegraphics[width =\textwidth]{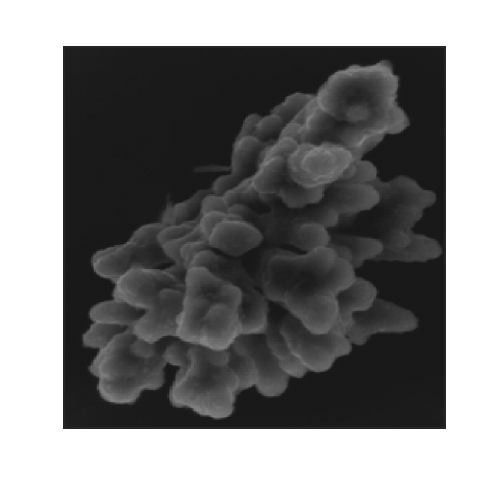}
        \scriptsize{\text{ }}
    \end{minipage}
    \begin{minipage}{     .17\linewidth}
        \centering
        \vspace{1mm}
        \text{\textbf{\normalsize $\etabaseline$}}
        \includegraphics[width =\textwidth]{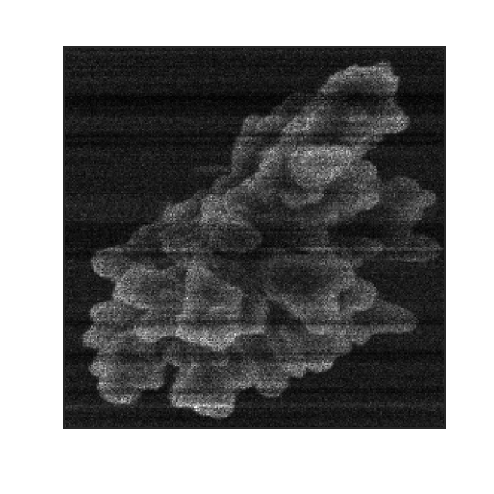}
        \scriptsize{\text{MSE = 0.5030}}
    \end{minipage}
    \begin{minipage}{     .17\linewidth}
        \centering
        \vspace{1mm}
        \text{\textbf{\normalsize $\etaft$}}
        \includegraphics[width =\textwidth]{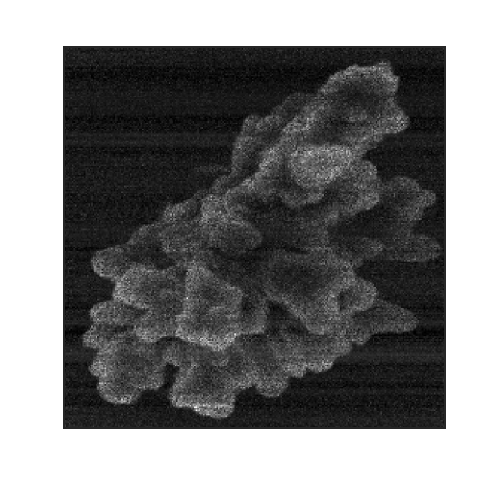}
        \scriptsize{\text{MSE = 0.3876}}
    \end{minipage}
      \begin{minipage}{     .17\linewidth}
        \centering
        \vspace{1mm}
        \text{\textbf{\normalsize $\lambda$}}
        \includegraphics[width =\textwidth]{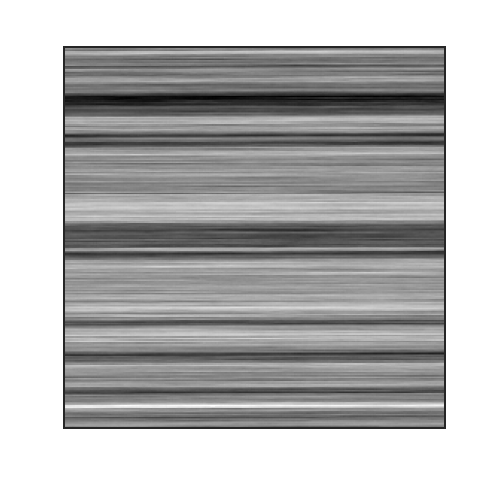}
        \scriptsize{\text{ }}
    \end{minipage}
    \begin{minipage}{     .17\linewidth}
        \centering
        \vspace{1mm}
        \text{\textbf{\normalsize $\lamref$}}
        \includegraphics[width =\textwidth]{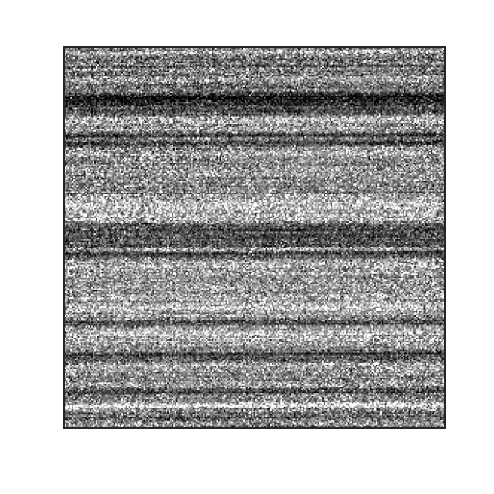}
        \scriptsize{\text{MSE = 33.5090}}
    \end{minipage}
    \\
    \begin{minipage}{     .17\linewidth}
        \centering
        \vspace{1mm}
        \text{\textbf{\normalsize $\etatrml$}}
        \includegraphics[width =\textwidth]{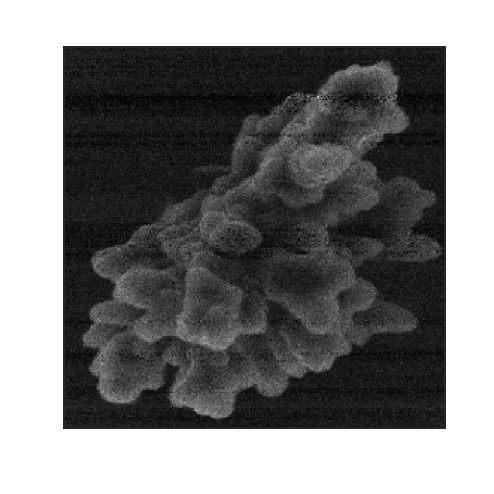}
        \scriptsize{\text{MSE = 0.1806}}
    \end{minipage}
    \begin{minipage}{     .17\linewidth}
        \centering
        \vspace{1mm}
        \text{\textbf{\normalsize $\etaseq$}}
        \includegraphics[width =\textwidth]{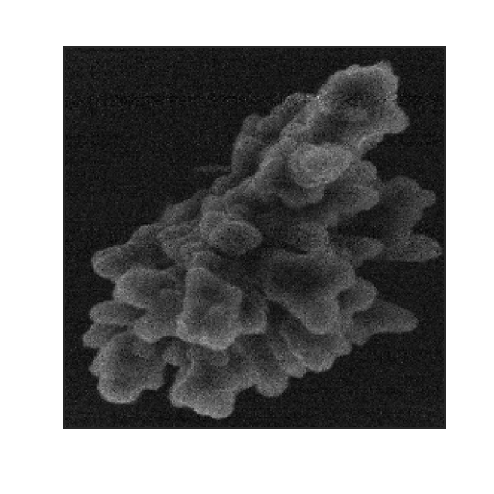}
        \scriptsize{\text{MSE = 0.1581}}
    \end{minipage}
    \begin{minipage}{     .17\linewidth}
        \centering
        \vspace{1mm}
        \text{\textbf{\normalsize $\etaoracle$}}
        \includegraphics[width =\textwidth]{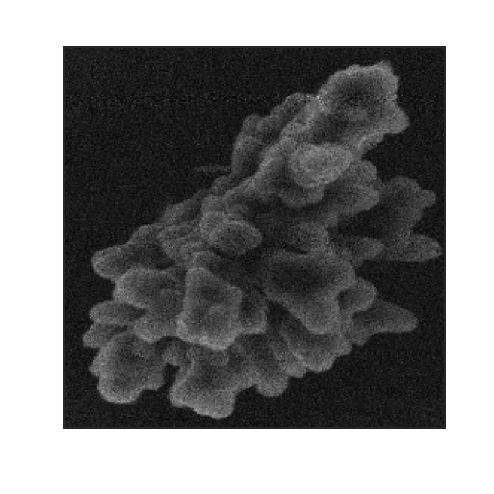}
        \scriptsize{\text{MSE = 0.1559}}
    \end{minipage}
    \begin{minipage}{     .17\linewidth}
        \centering
        \vspace{1mm}
        \text{\textbf{\normalsize $\lamseqTWO$}}
        \includegraphics[width =\textwidth]{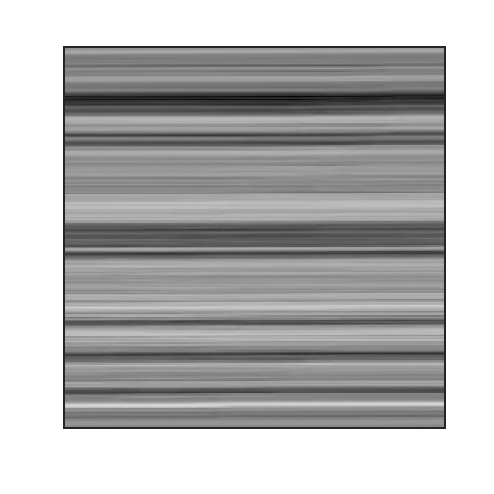}
        \scriptsize{\text{MSE = 2.2309}}
    \end{minipage}
    \begin{minipage}{     .17\linewidth}
        \centering
        \vspace{1mm}
        \text{\textbf{\normalsize $\lamseqONE$}}
        \includegraphics[width =\textwidth]{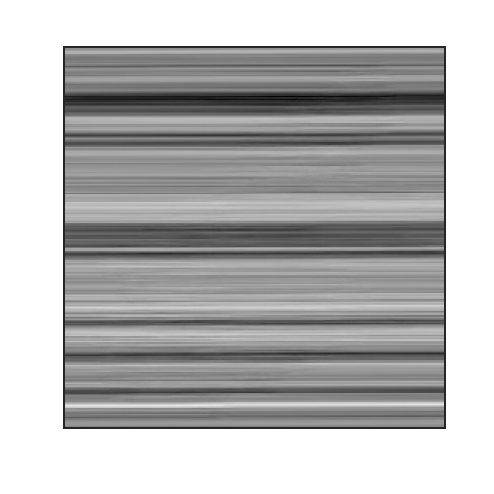}
        \scriptsize{\text{MSE = 1.7675}}
    \end{minipage}
    \\
    \begin{minipage}{1\textwidth}
        \vspace{1mm}
        \text{\textbf{Example 2}: HIM-\emph{medium} with $\eta \in [1,\,5]$, $\tilde{\lambda} = 20$, $\sigma_{\lambda}/\tilde{\lambda} = 0.3$, and $a = 0.999$.}
    \end{minipage}
    \begin{minipage}{     .17\linewidth}
        \centering
        \includegraphics[width =\textwidth]{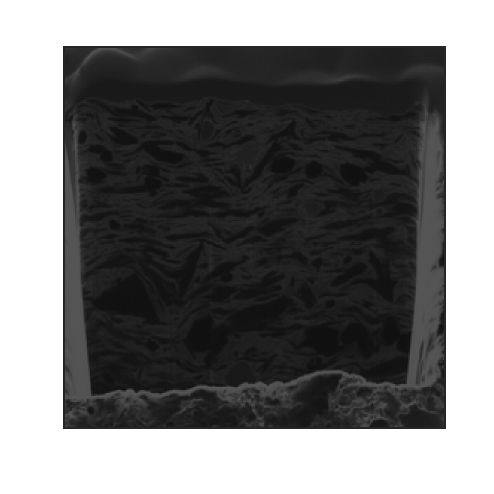}
        \scriptsize{\text{ }}
    \end{minipage}
    \begin{minipage}{     .17\linewidth}
        \centering
        \includegraphics[width =\textwidth]{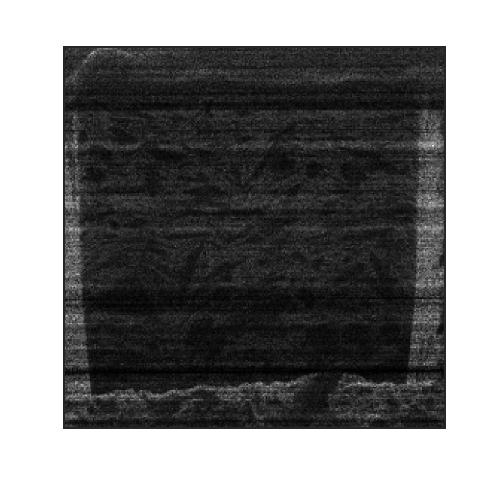}
        \scriptsize{\text{MSE = 0.7988}}
    \end{minipage}
    \begin{minipage}{     .17\linewidth}
        \centering
        \includegraphics[width =\textwidth]{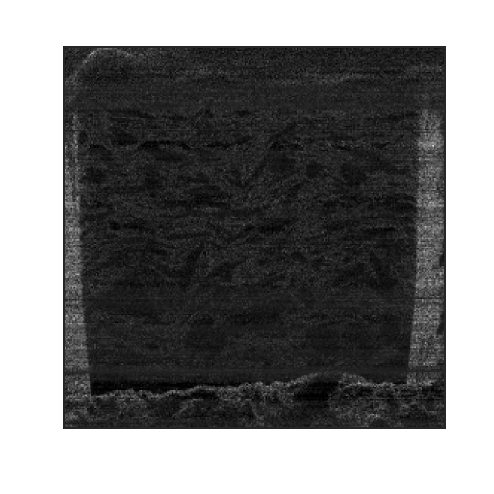}
        \scriptsize{\text{MSE = 0.5899}}
    \end{minipage}
    \begin{minipage}{     .17\linewidth}
        \centering
        \includegraphics[width =\textwidth]{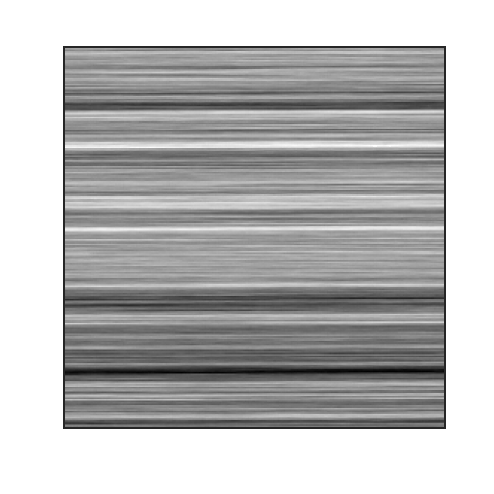}
        \scriptsize{\text{ }}
    \end{minipage}
    \begin{minipage}{     .17\linewidth}
        \centering
        \includegraphics[width =\textwidth]{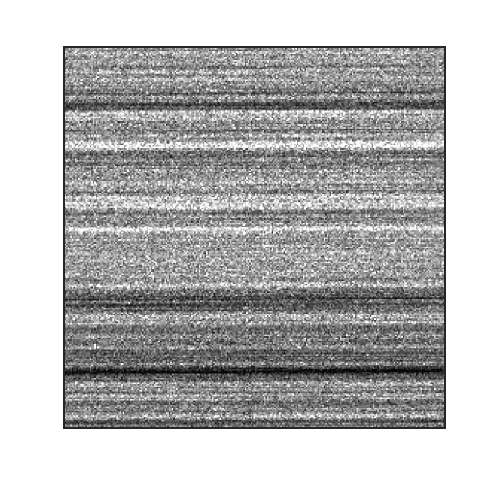}
        \scriptsize{\text{MSE = 32.4821}}
    \end{minipage}
    \\
    \begin{minipage}{     .17\linewidth}
        \centering
        \includegraphics[width =\textwidth]{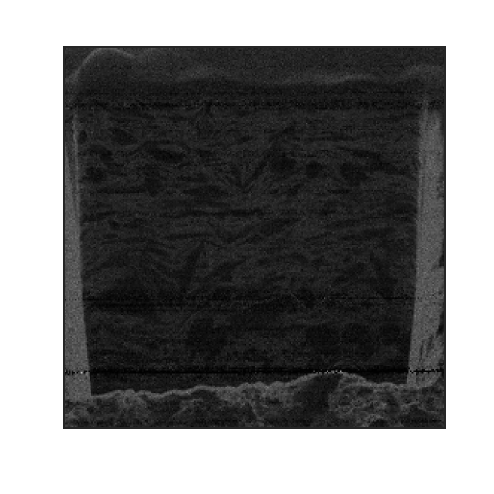}
        \scriptsize{\text{MSE = 0.2231}}
    \end{minipage}
    \begin{minipage}{     .17\linewidth}
        \centering
        \includegraphics[width =\textwidth]{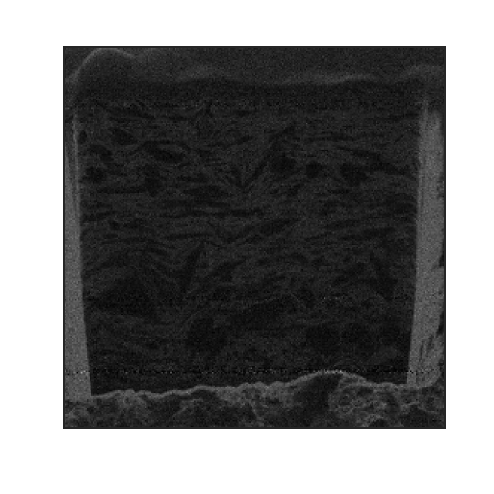}
        \scriptsize{\text{MSE = 0.1838}}
    \end{minipage}
     \begin{minipage}{     .17\linewidth}
        \centering
        \includegraphics[width =\textwidth]{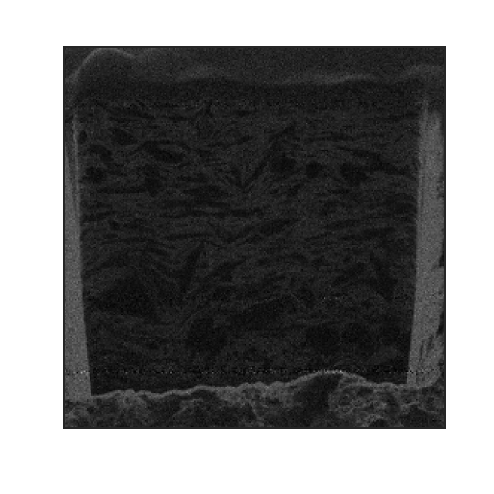}
        \scriptsize{\text{MSE = 0.1807}}
    \end{minipage}
    \begin{minipage}{     .17\linewidth}
        \centering
        \includegraphics[width =\textwidth]{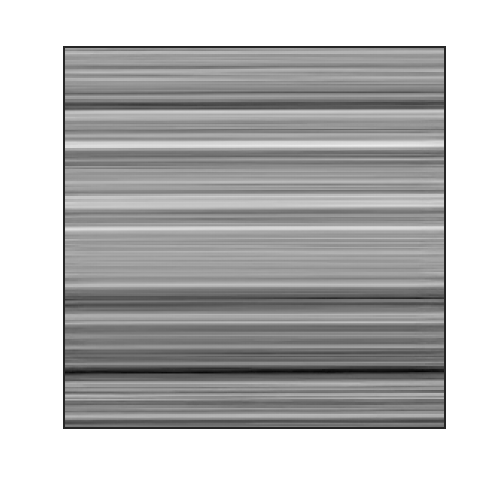}
        \scriptsize{\text{MSE = 2.8157}}
    \end{minipage}
    \begin{minipage}{     .17\linewidth}
        \centering
        \includegraphics[width =\textwidth]{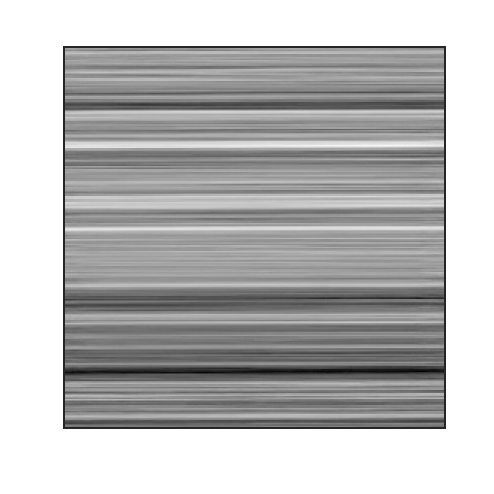}
        \scriptsize{\text{MSE = 2.0301}}
    \end{minipage}
    \\
    \begin{minipage}{1\textwidth}
        \vspace{1mm}
        \text{\textbf{Example 3}: HIM-\emph{slow} with $\eta \in [1,\,5]$,
     $\tilde{\lambda} = 20$, $\sigma_{\lambda}/\tilde{\lambda} = 0.2$, and $a = 0.9999$.}
    \end{minipage}
    \begin{minipage}{     .17\linewidth}
        \centering
        \includegraphics[width =\textwidth]{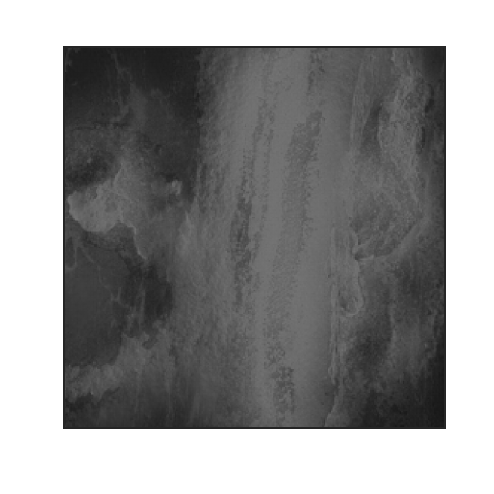}
        \scriptsize{\text{ }}
    \end{minipage}
    \begin{minipage}{     .17\linewidth}
        \centering
        \includegraphics[width =\textwidth]{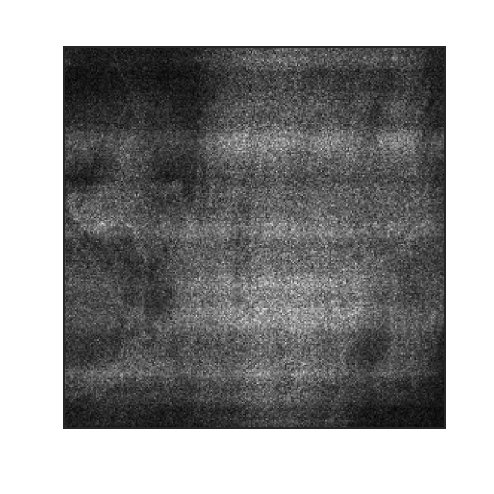}
        \scriptsize{\text{MSE = 1.2523}}
    \end{minipage}
    \begin{minipage}{     .17\linewidth}
        \centering
        \includegraphics[width =\textwidth]{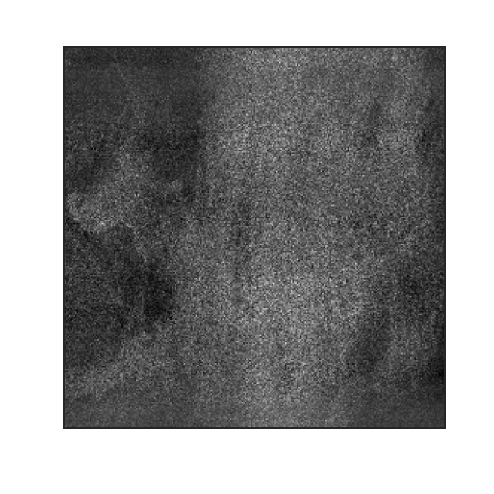}
        \scriptsize{\text{MSE = 0.9883}}
    \end{minipage}
    \begin{minipage}{     .17\linewidth}
        \centering
        \includegraphics[width =\textwidth]{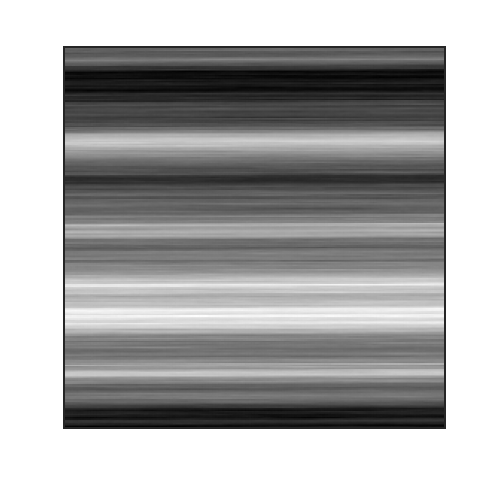}
        \scriptsize{\text{ }}
    \end{minipage}
    \begin{minipage}{     .17\linewidth}
        \centering
        \includegraphics[width =\textwidth]{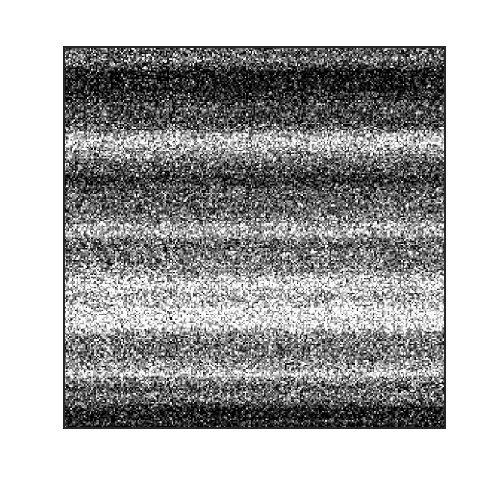}
        \scriptsize{\text{MSE = 26.0650}}
    \end{minipage}
    \\
    \begin{minipage}{     .17\linewidth}
        \centering
        \includegraphics[width =\textwidth]{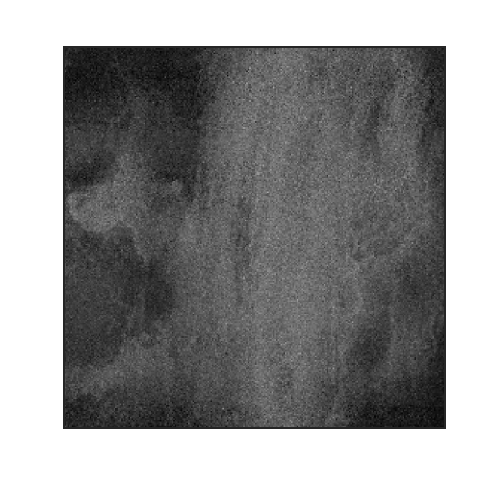}
        \scriptsize{\text{MSE = 0.2680}}
    \end{minipage}
    \begin{minipage}{     .17\linewidth}
        \centering
        \includegraphics[width =\textwidth]{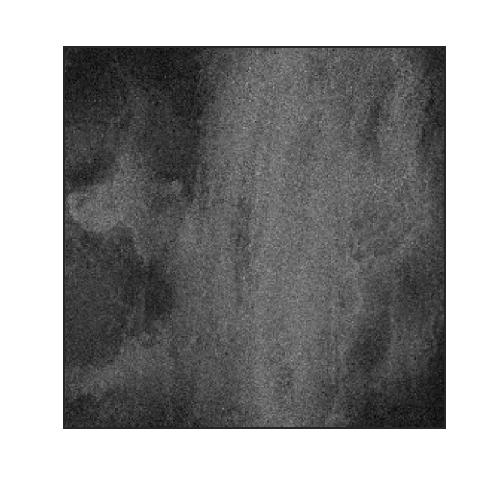}
        \scriptsize{\text{MSE = 0.2523}}
    \end{minipage}
    \begin{minipage}{     .17\linewidth}
        \centering
        \includegraphics[width =\textwidth]{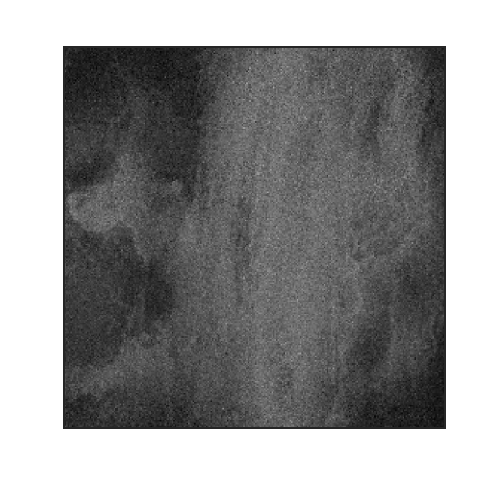}
        \scriptsize{\text{MSE = 0.2514}}
    \end{minipage}
    \begin{minipage}{     .17\linewidth}
        \centering
        \includegraphics[width =\textwidth]{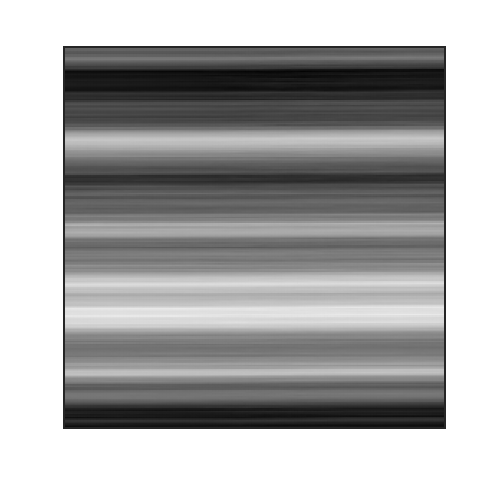}
        \scriptsize{\text{MSE = 0.3824}}
    \end{minipage}
    \begin{minipage}{     .17\linewidth}
        \centering
        \includegraphics[width =\textwidth]{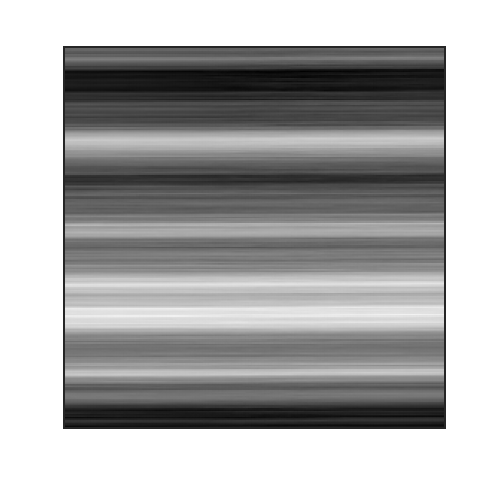}
        \scriptsize{\text{MSE = 0.2893}}
    \end{minipage}
    \caption{Comparison of $\eta$ and $\lambda$ estimators including $\etabaseline$, $\etaft$, $\etatrml$, $\etaseq$, $\etaoracle$, $\lamref$, $\lamseqTWO$, and $\lamseqONE$.}
    \label{fig:micrographs1}
\end{figure*}

\begin{figure*}
    \centering
    \begin{minipage}{1\textwidth}
        \text{\normalsize \textbf{Example 4}: HIM-\emph{fast} with $\eta \in [1,\,5]$, $\tilde{\lambda} = 20$, $\sigma_{\lambda}/\tilde{\lambda} = 0.2$, and $a = 0.9$.}
    \end{minipage}
    \begin{minipage}{     .17\linewidth}
        \centering
        \vspace{1mm}
        \text{\textbf{\normalsize $\eta$}}
        \includegraphics[width =\textwidth]{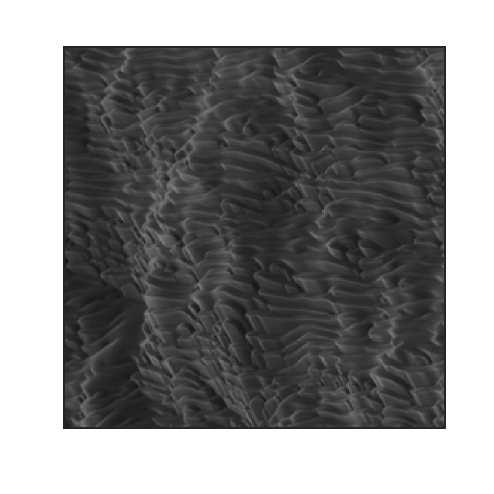}
        \scriptsize{\text{ }}
    \end{minipage}
    \begin{minipage}{     .17\linewidth}
        \centering
        \vspace{1mm}
        \text{\textbf{\normalsize $\etabaseline$}}
        \includegraphics[width =\textwidth]{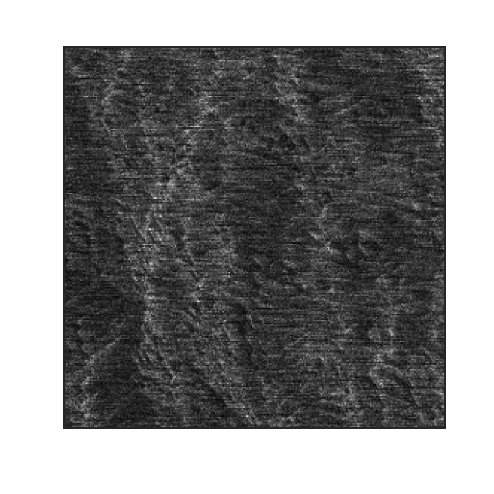}
        \scriptsize{\text{MSE = 0.6213}}
    \end{minipage}
    \begin{minipage}{     .17\linewidth}
        \centering
        \vspace{1mm}
        \text{\textbf{\normalsize $\etaft$}}
        \includegraphics[width =\textwidth]{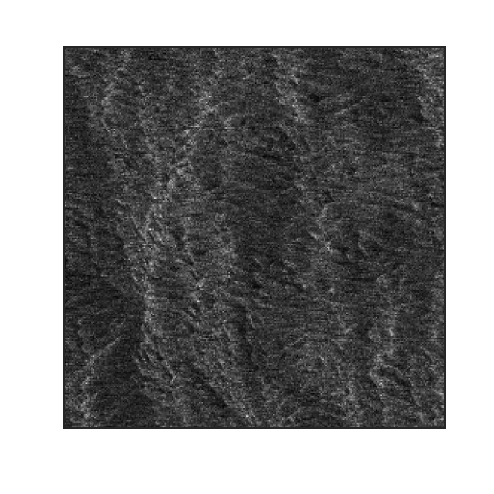}
        \scriptsize{\text{MSE = 0.5467}}
    \end{minipage}
    \begin{minipage}{     .17\linewidth}
        \centering
        \vspace{1mm}
        \text{\textbf{\normalsize $\lambda$}}
        \includegraphics[width =\textwidth]{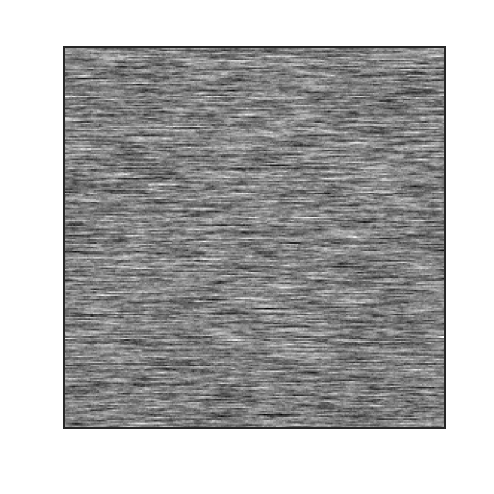}
        \scriptsize{\text{ }}
    \end{minipage}
    \begin{minipage}{     .17\linewidth}
        \centering
        \vspace{1mm}
        \text{\textbf{\normalsize $\lamref$}}
        \includegraphics[width =\textwidth]{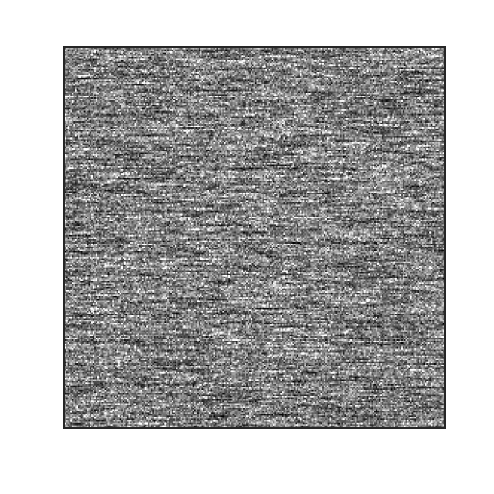}
        \scriptsize{\text{MSE = 29.1970}}
    \end{minipage}
    \\
    \begin{minipage}{     .17\linewidth}
        \centering
        \vspace{1mm}
        \text{\textbf{\normalsize $\etatrml$}}
        \includegraphics[width =\textwidth]{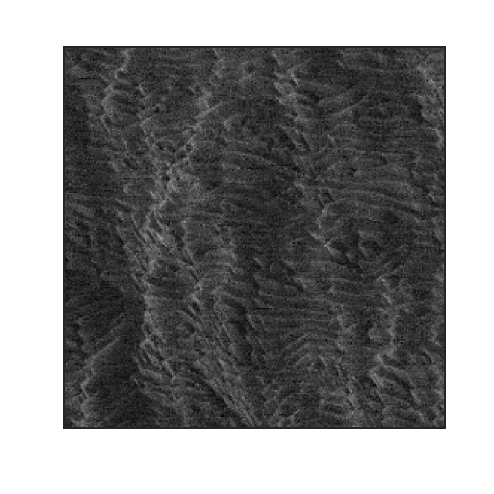}
        \scriptsize{\text{MSE = 0.2151}}
    \end{minipage}
    \begin{minipage}{     .17\linewidth}
        \centering
        \vspace{1mm}
        \text{\textbf{\normalsize $\etaseq$}}
        \includegraphics[width =\textwidth]{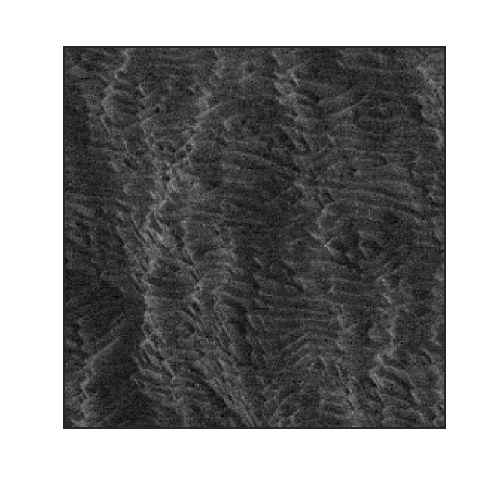}
        \scriptsize{\text{MSE = 0.2054}}
    \end{minipage}
    \begin{minipage}{     .17\linewidth}
        \centering
        \vspace{1mm}
        \text{\textbf{\normalsize $\etaoracle$}}
        \includegraphics[width =\textwidth]{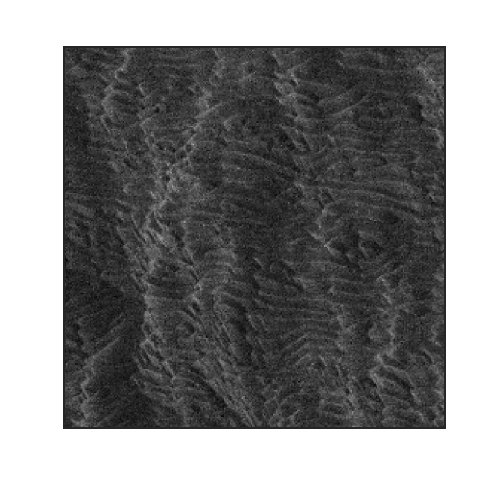}
        \scriptsize{\text{MSE = 0.1930}}
    \end{minipage}
    \begin{minipage}{     .17\linewidth}
        \centering
        \vspace{1mm}
        \text{\textbf{\normalsize $\lamseqTWO$}}
        \includegraphics[width =\textwidth]{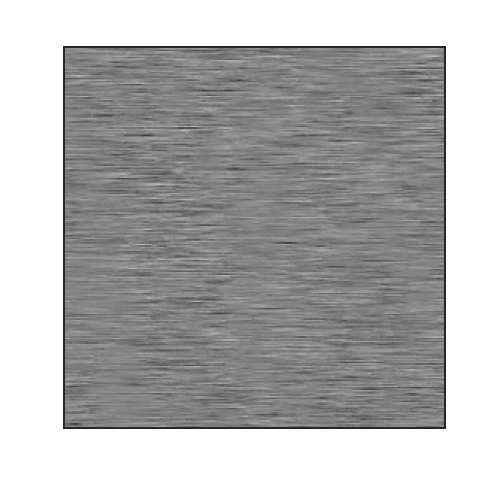}
        \scriptsize{\text{MSE = 8.4673}}
    \end{minipage}
    \begin{minipage}{     .17\linewidth}
        \centering
        \vspace{1mm}
        \text{\textbf{\normalsize $\lamseqONE$}}
        \includegraphics[width =\textwidth]{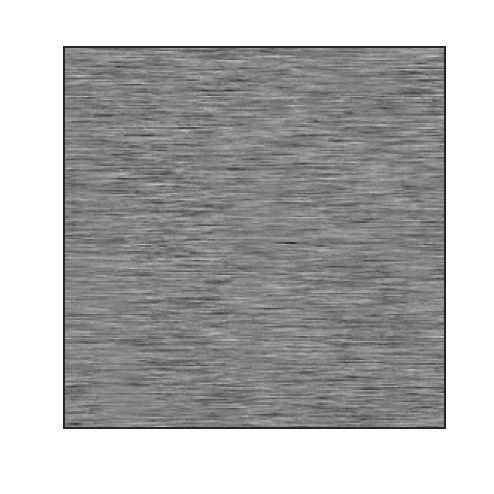}
        \scriptsize{\text{MSE = 7.3935}}
    \end{minipage}
    \\
    \begin{minipage}{1\textwidth}
        \vspace{1mm}
        \text{\normalsize \textbf{Example 5}: SEM-\emph{slow} with $\eta \in [0.2,\,1]$, $\tilde{\lambda} = 200$,$\sigma_{\lambda}/\tilde{\lambda} = 0.2$, and $a = 0.9999$.}
    \end{minipage}
    \begin{minipage}{     .17\linewidth}
        \centering
        \includegraphics[width =\textwidth]{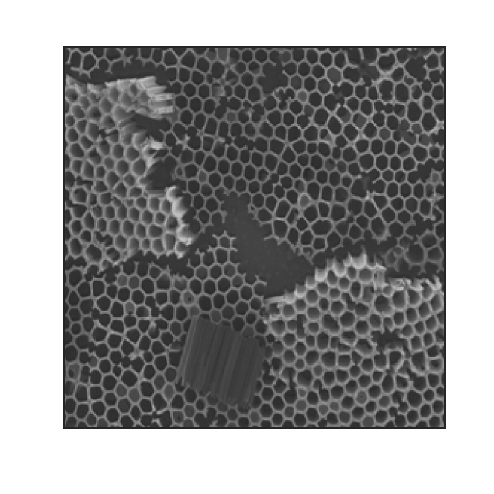}
        \scriptsize{\text{ }}
    \end{minipage}
    \begin{minipage}{     .17\linewidth}
        \centering
        \includegraphics[width =\textwidth]{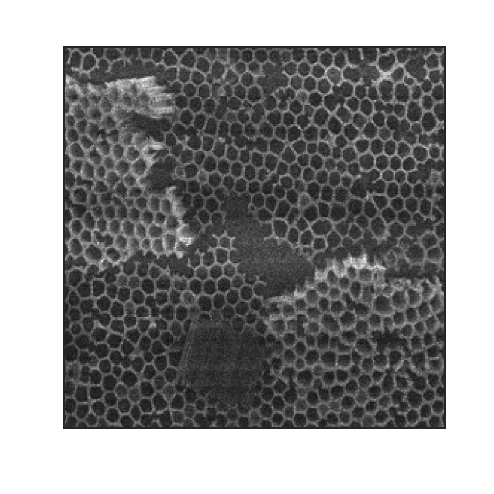}
        \scriptsize{\text{MSE = 0.0075}}
    \end{minipage}
    \begin{minipage}{     .17\linewidth}
        \centering
        \includegraphics[width =\textwidth]{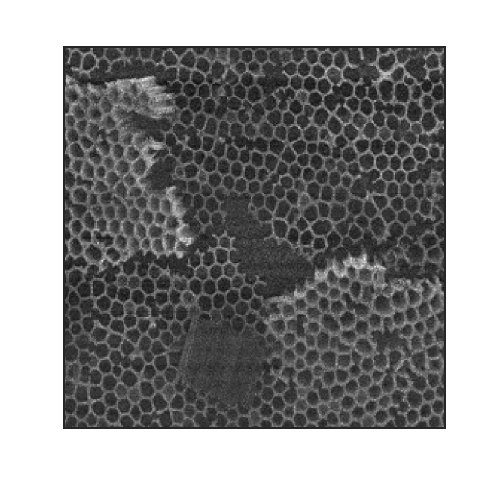}
        \scriptsize{\text{MSE = 0.0052}}
    \end{minipage}
    \begin{minipage}{     .17\linewidth}
        \centering
        \includegraphics[width =\textwidth]{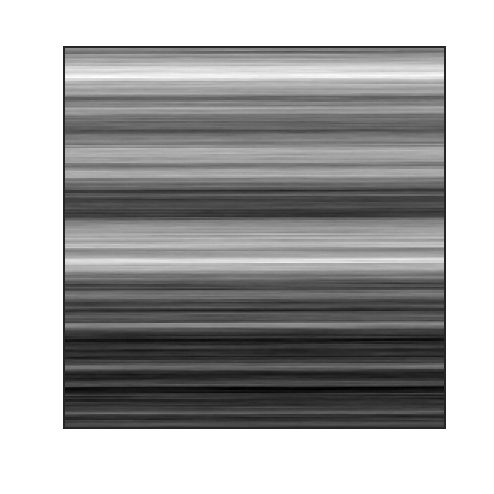}
        \scriptsize{\text{ }}
    \end{minipage}
    \begin{minipage}{     .17\linewidth}
        \centering
        \includegraphics[width =\textwidth]{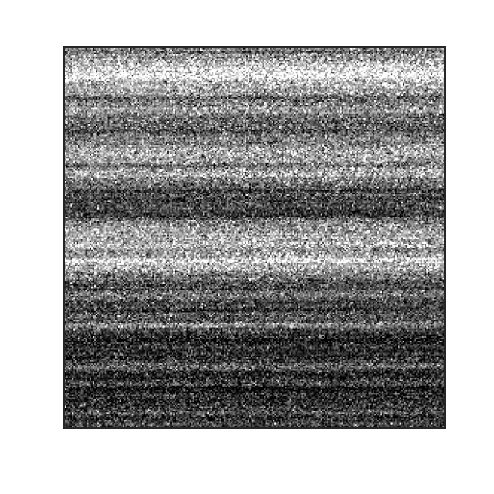}
        \scriptsize{\text{MSE = 680.3557}}
    \end{minipage}
    \\
    \begin{minipage}{     .17\linewidth}
        \centering
        \includegraphics[width =\textwidth]{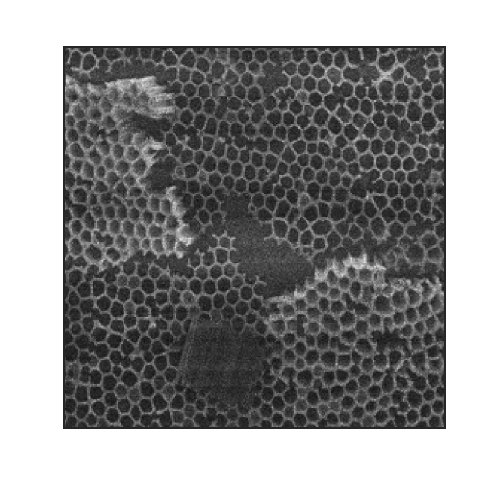}
        \scriptsize{\text{MSE = 0.0066}}
    \end{minipage}
    \begin{minipage}{     .17\linewidth}
        \centering
        \includegraphics[width =\textwidth]{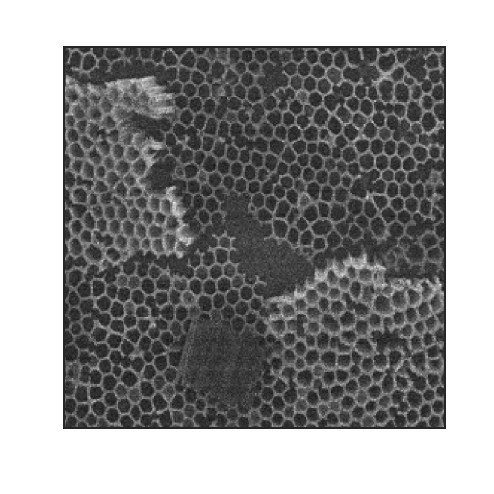}
        \scriptsize{\text{MSE = 0.0036}}
    \end{minipage}
    \begin{minipage}{     .17\linewidth}
        \centering
        \includegraphics[width =\textwidth]{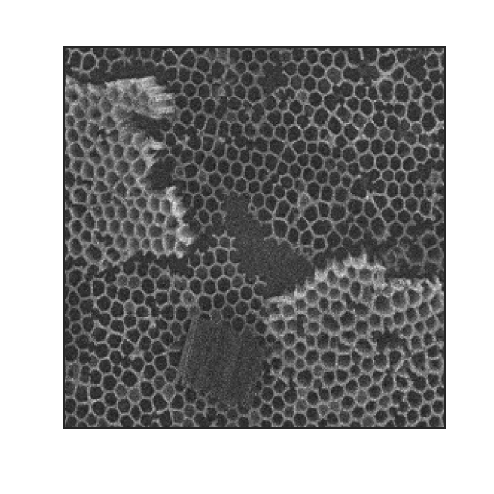}
        \scriptsize{\text{MSE = 0.0032}}
    \end{minipage}
     \begin{minipage}{     .17\linewidth}
        \centering
        \includegraphics[width =\textwidth]{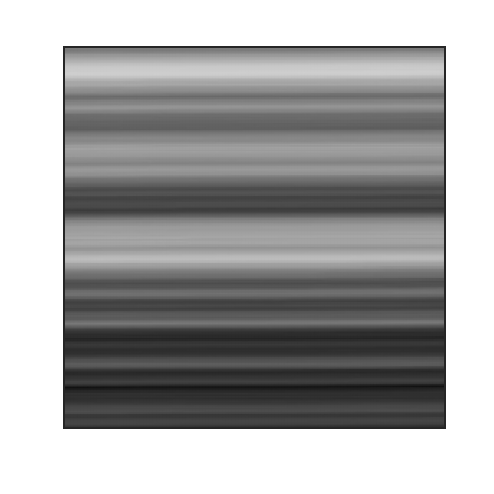}
        \scriptsize{\text{MSE = 91.6763}}
    \end{minipage}
    \begin{minipage}{     .17\linewidth}
        \centering
        \includegraphics[width =\textwidth]{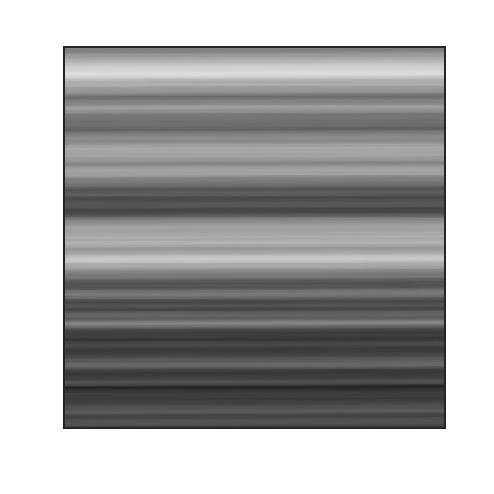}
        \scriptsize{\text{MSE = 87.3932}}
    \end{minipage}
    \\
    \begin{minipage}{1\textwidth}
        \vspace{1mm}
        \text{\normalsize \textbf{Example 6}: SEM-\emph{fast} with $\eta \in [0.2,\,1]$, $\tilde{\lambda} = 200$,$\sigma_{\lambda}/\tilde{\lambda} = 0.2$, and $a = 0.9$.}
    \end{minipage}
    \begin{minipage}{     .17\linewidth}
        \centering
        \includegraphics[width =\textwidth]{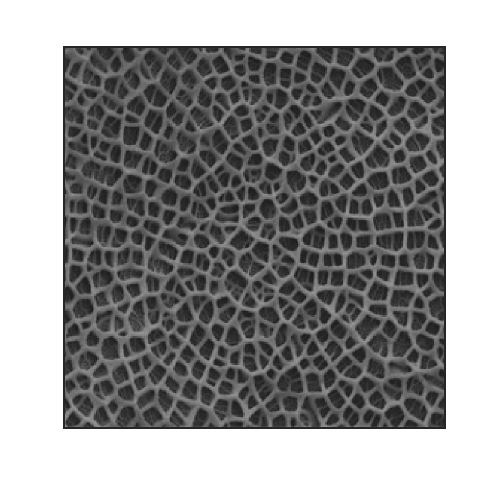}
        \scriptsize{\text{ }}
    \end{minipage}
    \begin{minipage}{     .17\linewidth}
        \centering
        \includegraphics[width =\textwidth]{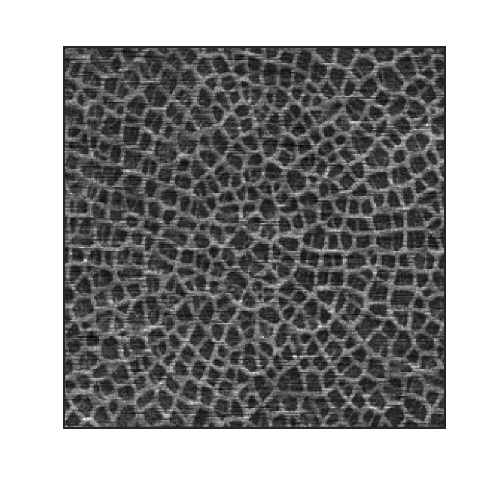}
        \scriptsize{\text{MSE = 0.0170}}
    \end{minipage}
    \begin{minipage}{     .17\linewidth}
        \centering
        \includegraphics[width =\textwidth]{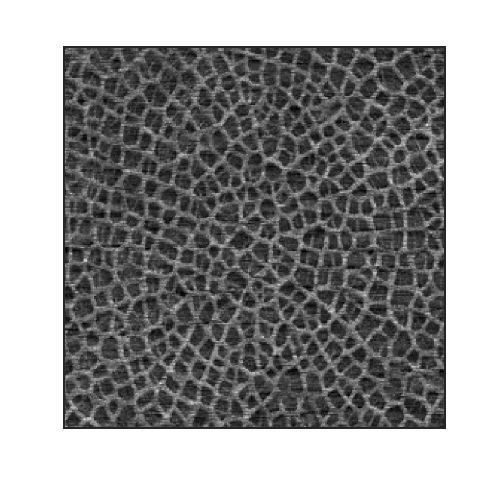}
        \scriptsize{\text{MSE = 0.0158}}
    \end{minipage}
    \begin{minipage}{     .17\linewidth}
        \centering
        \includegraphics[width =\textwidth]{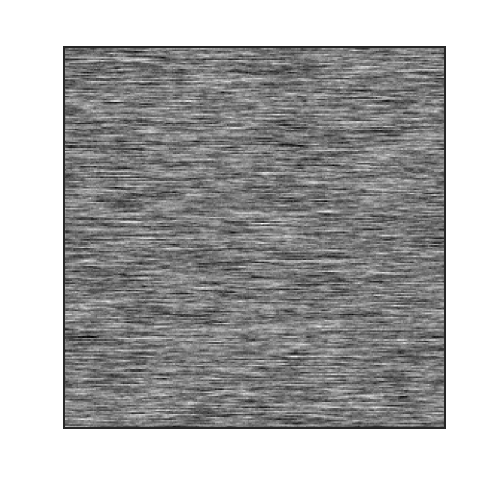}
        \scriptsize{\text{ }}
    \end{minipage}
    \begin{minipage}{     .17\linewidth}
        \centering
        \includegraphics[width =\textwidth]{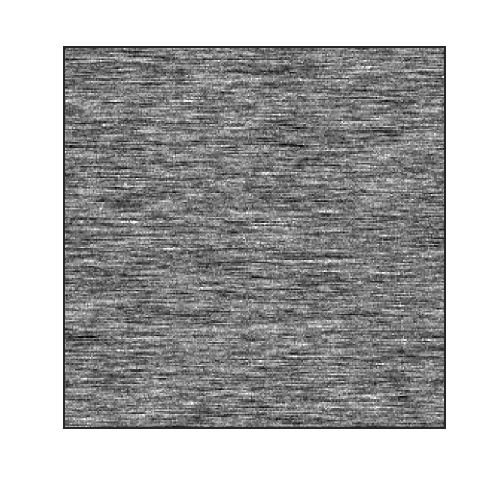}
        \scriptsize{\text{MSE = 659.9947}}
    \end{minipage}
    \\
    \begin{minipage}{     .17\linewidth}
        \centering
        \includegraphics[width =\textwidth]{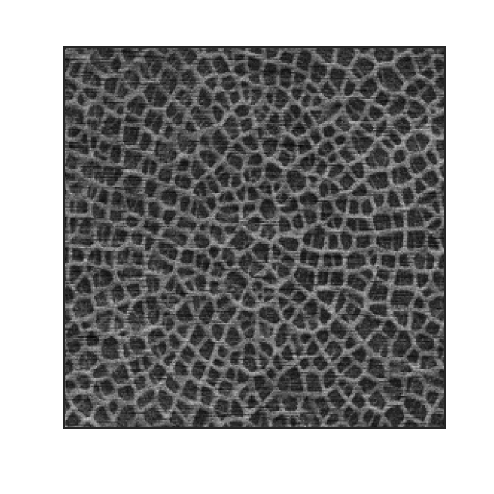}
        \scriptsize{\text{MSE = 0.0121}}
    \end{minipage}
    \begin{minipage}{     .17\linewidth}
        \centering
        \includegraphics[width =\textwidth]{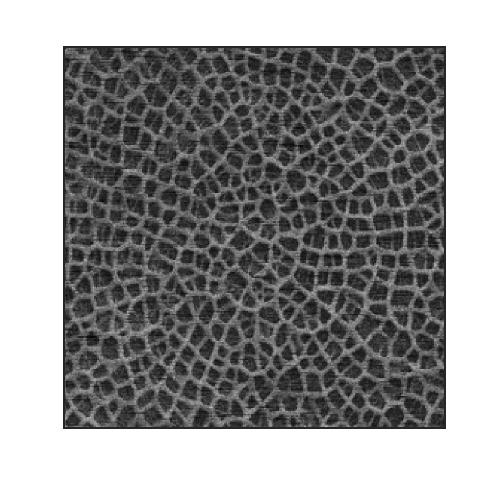}
        \scriptsize{\text{MSE = 0.0087}}
    \end{minipage}
    \begin{minipage}{     .17\linewidth}
        \centering
        \includegraphics[width =\textwidth]{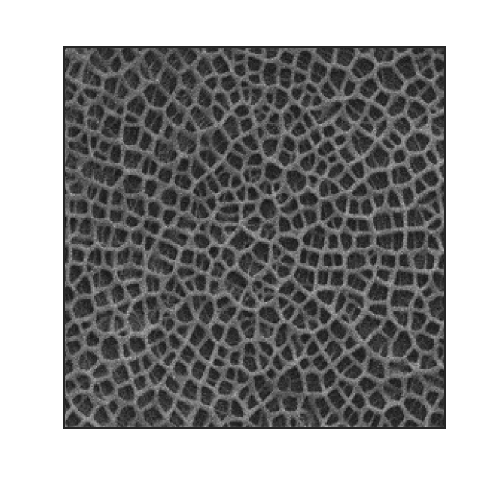}
        \scriptsize{\text{MSE = 0.0041}}
    \end{minipage}
    \begin{minipage}{     .17\linewidth}
        \centering
        \includegraphics[width =\textwidth]{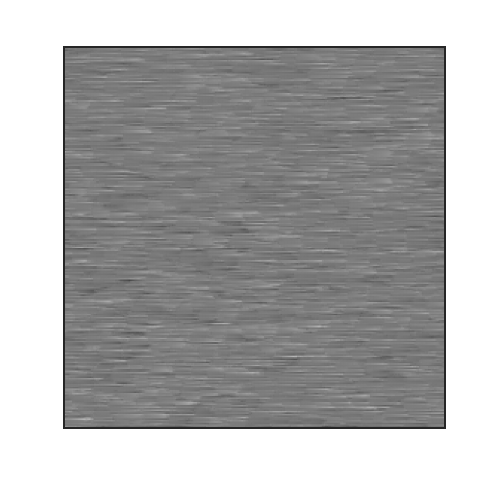}
        \scriptsize{\text{MSE = 1065.4336}}
    \end{minipage}
    \begin{minipage}{     .17\linewidth}
        \centering
        \includegraphics[width =\textwidth]{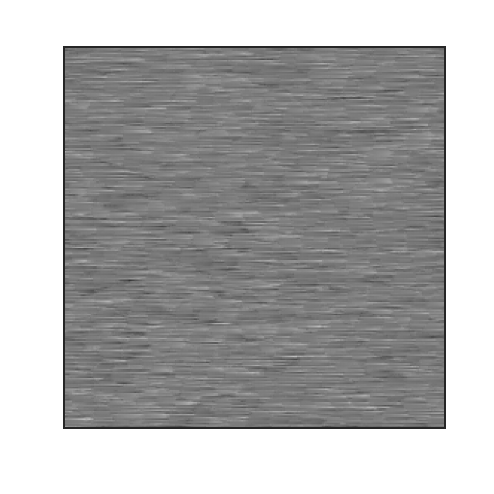}
        \scriptsize{\text{MSE = 943.6896}}
    \end{minipage}
    \caption{Comparison of $\eta$ and $\lambda$ estimators including $\etabaseline$, $\etaft$, $\etatrml$, $\etaseq$, $\etaoracle$, $\lamref$, $\lamseqTWO$, and $\lamseqONE$.}
    \label{fig:micrographs2}
\end{figure*}

\begin{table*}
 \begin{center}
  \begin{tabular}{|r|rrrl|rrrr|rrrr|}
\hline
   & & & & & \multicolumn{4}{c|}{Excess MSE} 
   & \multicolumn{4}{c|}{Excess MSE \%} \\
   Example & $\eta$ range & $\lamtilde$ & $\sigma_\lambda/\lamtilde$ & $a$
   & $\etabaseline$ & $\etaft$ & $\etatrml$ & $\etaseq$ 
   & $\etabaseline$ & $\etaft$ & $\etatrml$ & $\etaseq$ \\ \hline
        1  & [1,5]      & 20        & 0.2       & 0.999
            & 0.3470    & 0.2317    & 0.0247    & 0.0022
            & 100       & 66.8      & 7.1       & 0.6   \\
            
        2  & [1,5]      & 20        & 0.3       & 0.999
            & 0.6181    & 0.4092    & 0.0424    & 0.0031
            & 100       & 66.2      & 6.9       & 0.5   \\
            
        3  & [1,5]      & 20        & 0.2       & 0.9999
            & 1.0009    & 0.7369    & 0.0166    & 0.0009
            & 100       & 73.6      & 1.7       & 0.1   \\
            
        4  & [1,5]      & 20        & 0.2       & 0.9
            & 0.4283    & 0.3537    & 0.0221    & 0.0124
            & 100       & 82.6      & 5.2       & 2.9   \\
            
        5  & [0.2,1]    & 200       & 0.2       & 0.9999
            & 0.0043    & 0.0020    & 0.0034    & 0.0004
            & 100       & 46.5      & 79.1      & 9.3   \\
            
        6  & [0.2,1]    & 200       & 0.2       & 0.9
            & 0.0129    & 0.0117    & 0.0080    & 0.0046
            & 100       & 90.7      & 62.0      & 35.7  \\
        \hline
  \end{tabular}
 \end{center}
 \caption{Comparison of $\eta$ estimators for the examples simulated in Figures~\ref{fig:micrographs1} and~\ref{fig:micrographs2}. Excess MSE refers to the MSE beyond that of $\etaoracle$, which is the absolute best performance for these estimators. Excess MSE \% refers to the percentage of the excess MSE of $\etabaseline$ that is retained by using the different estimators.}
 \label{table:excess}
\end{table*}

\subsection{Comparisons for Six Different Scenarios}
The first of these is $\etabaseline$ (\ref{eq:conv}), which uses the constant incorrect dose value, $\lamtilde$. The next is $\etaft$, which appears to be the state of the art and is a \emph{DFT-domain nulling} algorithm that uses a 2D Fast Fourier Transform to take $\etabaseline$, transforms it into the Fourier domain, filters out the stripe contributions, and transforms it out of the Fourier domain~\cite{barlow2016removing,barlow2018observing}. More specifically, this algorithm tries to improve $\etabaseline$ and remove its stripe artifacts by filtering out the coefficients that correspond to those stripes. Once out of the Fourier domain, $\etaft$ is simply a cleaned up version of $\etabaseline$. In each application of $\etaft$, the coefficients that are nulled minimize the MSE, even though this would not be practically possible. In order to do this, filter parameters $w$ and $h$ are manually selected  to yield optimal performance of $\etaft$. Then in the frequency domain, where $k$ and $u$ are the horizontal and vertical frequency indices, respectively, the coefficients that satisfy both $|k| \leq w$ and $|u| > h$ are nulled. Lastly, the inverse transform is applied, giving the resulting image.
Next, there is $\etatrml$ (\ref{eq:trml}), which \emph{neglects} the current variation and uses $\lamtilde$, but proves to still be robust to this variation. The last method the new algorithm compares against is $\etaoracle$, which is $\etatrml$ (\ref{eq:trml}) with \emph{perfect} knowledge of the current variation and uses $\lambda$. 

Although $\etaoracle$ is obviously unattainable, it allows us to have a goal with $\etaseq$: to get within 20\% of the performance of $\etaoracle$ from $\etabaseline$. The design objective of the $\lambda$ estimation is less straightforward since this is more of a novelty; dose estimation is not typically attempted in PBM\@. Therefore, the goal of $\lamseqONE$, which is the algorithm's output dose estimate, is to at least improve upon $\lamref$ (\ref{eq:lam_ref}).

In both Figures~\ref{fig:micrographs1} and~\ref{fig:micrographs2} we have our simulation results and comparisons. Synthetic data was produced using existing micrographs \cite{Image-Bouquet,Image-MilledShaleSample,Image-Thegoldisland,Image-Nanotexture,Image-Honeycombs,Image-Diatome} as ground truth images.
In total there are 6 different scenarios with different ranges of $\eta$ to signify realistic conditions for HIM or SEM, different $\lamtilde$ values, different amounts of current variation, and different speeds of current variation. The first four examples have a larger range of $\eta$, representing HIM~\cite{notte2007introduction}, whereas the last two examples have a smaller range, representing SEM, neglecting topographical effects~\cite{LinJ:05}. The $\lamtilde$ values are chosen to correspond to the specific $\eta$ range. The other two parameters altered are the amount and speed of variation. All of the simulated examples were tested with nominal sub-acquisition dose $\tilde{\lambda}/n = 0.1$.

Increasing the amount of variation, or the coefficient of variation $\sigma_{\lambda}/\lamtilde$ of $\lambda$, makes it harder to estimate $\lambda$ since it varies more from its mean $\lamtilde$; this effect on the performance of these estimators is explored in Example 2. In terms of the speed of variation, this is thought of in terms of the correlation coefficient $a$ (\ref{eq:model}), which determines how closely correlated neighboring pixel's values of $\lambda$ are. When increasing $a$, there is more correlation between $\lambda$ values, meaning it varies more slowly over the pixels, making the striping artifacts extra thick. This can break down $\etaft$ since this would make it want to try to correct huge pieces of the image, which could end very badly. When decreasing $a$, there is less correlation between $\lambda$ values, meaning it varies more quickly over the pixels, making the striping not even across the entire width of the image, but lots of striping sections even across the image in one line. This can break down $\etaft$ since it is designed to remove horizontal stripes but with this variation, the little striping sections could be confused to be part of the image itself instead of being artifacts. In Examples 1 and 2, $a$ is chosen so the speed of variation is medium, whereas Examples 3 and 5 have slow variation and Examples 4 and 6 have fast variation.

As seen in all of these examples, although $\etaft$ is able to improve upon $\etabaseline$, $\etatrml$ proves to be extremely robust to this particle beam current variation. It is only in Example~5, which is an SEM image with slow variation, that $\etaft$ is able to outperform $\etatrml$; this is due to the fact that low values of $\eta$ make it extremely hard to estimate $\eta$ and in this case, the thick stripes due to slow current variation work in favor of $\etaft$. In terms of the new algorithm, in all of the examples simulated and tested, $\etaseq$ is successful at achieving the previously stated goal. In all examples, this algorithm manages to get remarkably close to $\etaoracle$, which has \emph{perfect} knowledge of the dose, demonstrating just how important the dose is to estimating the image. The algorithm also estimates the dose $\lambda$ tremendously, whether initialized with the correct $\lamtilde$ or not. Out of all the examples, it is only in Example~6 that this algorithm performs worse than $\lamref$; this demonstrates how difficult it is to estimate the dose when $\eta$ is small and current variation is fast.

To understand the results of $\eta$ estimation more easily, we can use the data in Table~\ref{table:excess}. In this table, each of the examples that were simulated in Figures~\ref{fig:micrographs1} and~\ref{fig:micrographs2} are explored further, specifically looking at the excess MSE and the percentage of it left for the various estimators. Excess MSE can be thought of as
\begin{equation}
    {\rm MSE}_{\rm excess}\big(\etahat\big) = {\rm MSE}\big(\etahat\big) - {\rm MSE}\big(\etaoracle\big).
\end{equation}
We can see that for the first four examples that represent HIM, a small percentage of the excess MSE due to beam current variation is retained with $\etatrml$, much less than that of $\etaft$, illustrating the immense capability of $\etatrml$ on its own. Estimating the beam current with $\etaseq$ improves upon this even further, leaving only a very small percentage of the excess MSE\@. As mentioned in the previous paragraph, $\etaft$ outperforms $\etatrml$ in Example 5, which can be seen more clearly by inspecting the excess MSE~\% in this table, although $\etaseq$ improves upon both of these estimators, with a much smaller excess MSE~\% remaining. For the last example representing SEM, once again a smaller percentage of the excess MSE is retained by estimating with $\etatrml$ compared to $\etaft$, and this is lowered even further by using $\etaseq$.

While the whole table demonstrates remarkable results for this alternating algorithm, only the last column is needed to be convinced of how close the results are for $\etaseq$ and $\etaoracle$, in terms of MSE\@. For all of the images, the percentage of the MSE for $\etaseq$ beyond that of $\etaoracle$ is extremely low, with the largest value being 35.7\%, which is much lower than those of the other estimators for that specific example.

The MSE of the algorithm's image estimates are the lowest of the attainable methods across all these simulated examples under different conditions. Unlike the DFT-domain nulling method, this algorithm, and also TRML itself, do not require inspecting a previously estimated image, meaning that they can be implemented even with naturally striped images without introducing any new artifacts.

\begin{figure*}
    \centering
    \begin{minipage}{1\textwidth}
        \text{\normalsize \textbf{Example 1-extended}: HIM-\emph{medium} with $\eta \in [1,\,5]$, $\tilde{\lambda} = 20$, $\sigma_{\lambda}/\tilde{\lambda} = 0.2$, and $a = 0.999$.}
    \end{minipage}
    \begin{minipage}{     .17\linewidth}
        \centering
        \vspace{1mm}
        \text{\textbf{\normalsize $\eta$}}
        \includegraphics[width =\textwidth]{ex1/truth.pdf}
        \scriptsize{\text{ }}
    \end{minipage}
    \begin{minipage}{     .17\linewidth}
        \centering
        \vspace{1mm}
        \text{\textbf{\normalsize $\etaseq$}}
        \includegraphics[width =\textwidth]{ex1/alg.pdf}
        \scriptsize{\text{MSE = 0.1581}}
    \end{minipage}
    \begin{minipage}{     .17\linewidth}
        \centering
        \vspace{1mm}
        \text{\textbf{\normalsize $\etaseqhigha$}}
        \includegraphics[width =\textwidth]{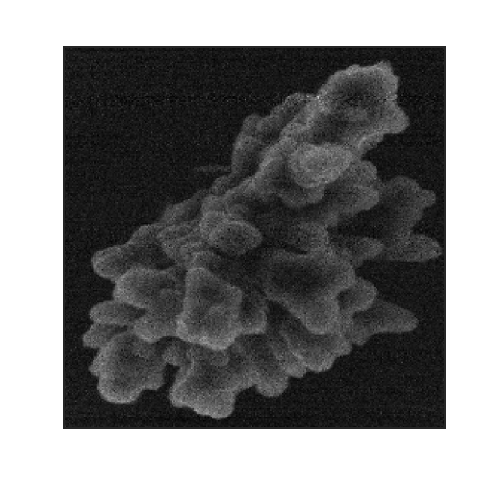}
        \scriptsize{\text{MSE = 0.1637}}
    \end{minipage}
    \begin{minipage}{     .17\linewidth}
        \centering
        \vspace{1mm}
        \text{\textbf{\normalsize $\etaseqlowa$}}
        \includegraphics[width =\textwidth]{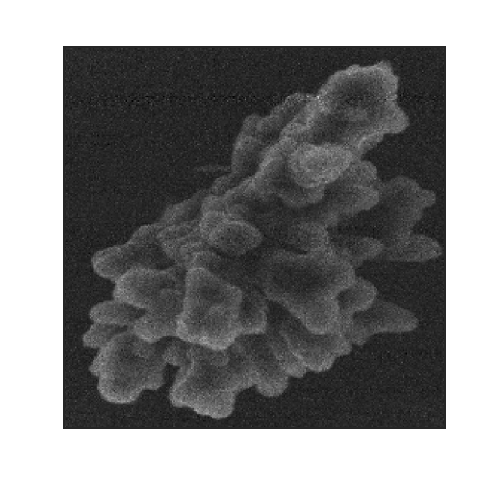}
        \scriptsize{\text{MSE = 0.3158}}
    \end{minipage}
    \begin{minipage}{     .17\linewidth}
        \centering
        \vspace{1mm}
        \text{\textbf{\normalsize $\etabaseline$}}
        \includegraphics[width =\textwidth]{ex1/conv.pdf}
        \scriptsize{\text{MSE = 0.5030}}
    \end{minipage}
    \\
    \begin{minipage}{     .17\linewidth}
        \centering
        \vspace{1mm}
        \text{\textbf{\normalsize $\lambda$}}
        \includegraphics[width =\textwidth]{ex1/ltruth.pdf}
        \scriptsize{\text{ }}
    \end{minipage}
    \begin{minipage}{     .17\linewidth}
        \centering
        \vspace{1mm}
        \text{\textbf{\normalsize $\lamseqONE$}}
        \includegraphics[width =\textwidth]{ex1/lalg1.pdf}
        \scriptsize{\text{MSE = 1.7675}}
    \end{minipage}
    \begin{minipage}{     .17\linewidth}
        \centering
        \vspace{1mm}
        \text{\textbf{\normalsize $\lamseqONEhigha$}}
        \includegraphics[width =\textwidth]{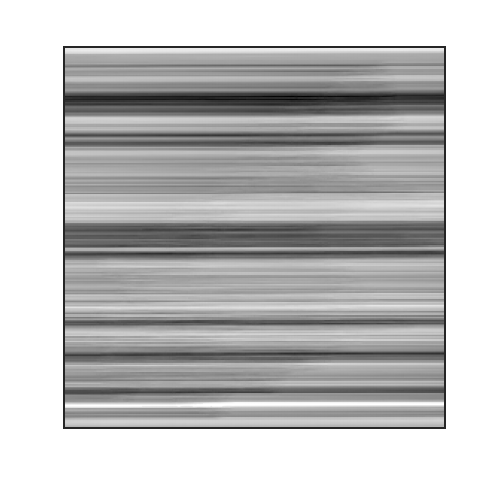}
        \scriptsize{\text{MSE = 8.1949}}
    \end{minipage}
    \begin{minipage}{     .17\linewidth}
        \centering
        \vspace{1mm}
        \text{\textbf{\normalsize $\lamseqONElowa$}}
        \includegraphics[width =\textwidth]{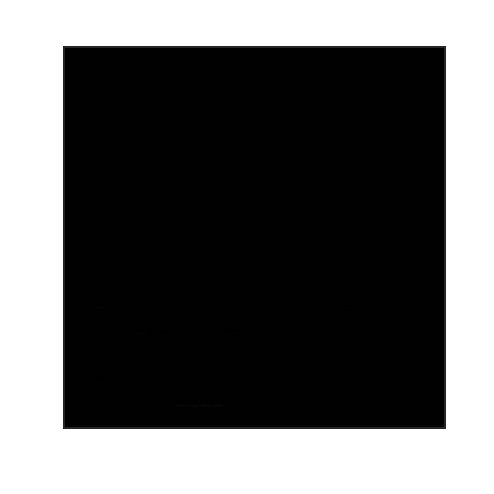}
        \scriptsize{\text{MSE = 271.1094}}
    \end{minipage}
    \begin{minipage}{     .17\linewidth}
        \centering
        \vspace{1mm}
        \text{\textbf{\normalsize $\lamref$}}
        \includegraphics[width =\textwidth]{ex1/loracle.pdf}
        \scriptsize{\text{MSE = 33.5090}}
    \end{minipage}
    \caption{Comparison of $\eta$ and $\lambda$ estimators including $\etaseq$, $\etaseqhigha$, $\etaseqlowa$, $\etabaseline$, $\lamseqONE$, $\lamseqONEhigha$, $\lamseqONElowa$, $\lamref$.}
    \label{fig:micographs3}
\end{figure*}

\subsection{Assuming Wrong Autocorrelation}
Lastly, in Figure~\ref{fig:micographs3}, Example 1 from Figure~\ref{fig:micrographs1} is revisited with more detail. In Figure~\ref{fig:micographs3}, all of the conditions from Example 1 are the same, but the wrong autocorrelation value $a$ is used when estimating. Specifically, this example has $a = 0.999$, but the alternating algorithm is run twice, once assuming $a = 0.9999$ and another assuming $a = 0.9$. 

When $a = 0.9999$ is assumed, this means the estimator, now $\etaseqhigha$ and $\lamseqONEhigha$, thinks that $\lambda$ varies slower than it actually is. With the assumption that there is more correlation between $\lambda$ values at different pixels, the estimator will want to keep the estimated dose at nearby pixels more similar to each other than they should be. As seen in the third column of Figure~\ref{fig:micographs3}, this does worsen the performance of the algorithm, but not extremely. Both $\etaseqhigha$ and $\lamseqONEhigha$ are able to outperform $\etabaseline$ and $\lamref$, respectively, although wrong assumptions are made.

When $a = 0.9$ is assumed, this means the estimator, now $\etaseqlowa$ and $\lamseqONElowa$, thinks that $\lambda$ varies faster than it actually is, causing issues. With the assumption that there is less correlation between $\lambda$ values at different pixels, the estimator will want the estimated dose at nearby pixels to differ from each other more than they should. As seen in the fourth column of Figure~\ref{fig:micographs3}, this worsens the performance of the algorithm much more than when $a$ was assumed to be too high. This is demonstrated especially with $\lamseqONElowa$, where $\lambda$ is assumed to vary much farther from the mean $\lamtilde$ than the reality, making the algorithm's first run of the sequential filter give a very low $\lambda$ estimate for the first pixel. Then acquiring more information by estimating $\eta$ helps the estimator, and the entire algorithm itself, stabilize, but at dose values much lower than the true $\lambda$. This results in a dark image for $\lamseqONElowa$ and a very high MSE\@. Although $\lamseqONElowa$ does not outperform $\lamref$ since it does such a poor job, $\etaseqlowa$ is still able to outperform $\etabaseline$.

This experiment clearly shows just how valuable knowing the autocorrelation $a$ is for this sequential filter and the alternating algorithm to work. Although this parameter is important for computing the best estimates of $\eta$ and $\lambda$, this algorithm still manages to produce an acceptable image estimate of the sample, demonstrating the power of time-resolved measurement, even in this scenario.

\section{Conclusion}
Quite recently in~\cite{PENG2020}, time-resolved measurement has been proven to achieve greater image quality than conventional measurement in the scenario of assumed known beam current. Now that this assumption of perfectly known beam current is taken away, I have been able to show how extremely robust time-resolved associated methods, specifically TRML, are to this beam current variation. Taking a step further, by estimating the dose with this new proposed algorithm, the limits of image estimation can be approached like never before, getting exceptionally close to what could be done with perfect knowledge of this varying dose. In addition, by estimating the dose, which is a novel idea, manufacturing and fabrication processes can be more reliable. Both of these results mean that the very precise and expensive particle beam generators previously used are not needed, substantially lowering the costs in labs, making PBM more accessible and manageable to all.

\appendix
This is the derivation of the sequential filter (\ref{eq:filter}) used in the alternating algorithm, which is reprinted here for convenience. 
\begin{equation}
    \label{eq:filter2}
    \lamhat_p(y_p) = \frac{\covrm(\lambda_p,y_p)}{\varrm(y_p)}(y_p - E[y_p]) + E[\lambda_p].
\end{equation}
To begin, we have our AR model of $\lambda$ (\ref{eq:model}), which has mean
\begin{align}
    E[\lambda_p] &= E[x_p + a\lambda_{p-1} + c] \nonumber \\
                 &= aE[\lambda_{p-1}] + c.
\end{align}
Since $E[\lambda_{p-1}]$ is assumed to equal $E[\lambda_p]$, this can be rearranged to give
\begin{equation}
\label{eq:armean}
E[\lambda_p] = \frac{c}{1 - a} = \frac{\Tilde{\lambda}(1 - a)}{1 - a} = \lamtilde,
\end{equation}
which is why $c = \lamtilde(1-a)$, in order to have the mean be $\lamtilde$, as mentioned in Section~\ref{section:modeldescription}.
The model (\ref{eq:model}) has variance
\begin{align}
    \varrm(\lambda_p) &= \varrm(x_p + a\lambda_{p-1} + c) \nonumber \\
                   &= \sigma^2_x + a^2\varrm(\lambda_{p-1}).
\end{align}
Similarly to the mean, $\varrm(\lambda_{p-1})$ is assumed to equal $\varrm(\lambda_p)$, meaning
\begin{equation}
\label{eq:arvar}
    \varrm(\lambda_p)= \frac{\sigma^2_x}{1 - a^2}.
\end{equation}

By looking at the sequential filter, we see that there are multiple terms that are needed. The first two have already been computed~\cite{PENG2020} and are the mean
\begin{equation}
\label{eq:basicymean}
    E[y_p|\eta_p,\lambda_p] = \lambda_p\eta_p,
\end{equation}
and the variance
\begin{equation}
\label{eq:basicyvar}
    \varrm(y_p|\eta_p,\lambda_p) = \lambda_p(\eta_p + \eta_p^2)
\end{equation}
of the secondary electrons $Y$, although these assume that both $\lambda$ and $\eta$ are somehow known at this specific pixel. Therefore, to have more accuracy, we can remove these assumptions, which means that the values of $\lambda$ and $\eta$ used are estimates and not exactly their true values. More specifically, we have $\etahat_p = \eta_p + \gamma_p$ and $\lamhat_{p-1} = \lambda_{p-1} + \epsilon_p$, where  $\epsilon_p \sim N(0,\sigeps)$ and $\gamma_p \sim N(0,\siggams)$.

The first term we need in (\ref{eq:filter2}) is $E[\lambda_p]$. Using (\ref{eq:model}),
\begin{align}
    E[\lambda_p|\hat\eta_p,\hat\lambda_{p-1}] &= E[\lambda_p|\hat\lambda_{p-1}] \nonumber \\
    &= E[x_p + a\lambda_{p-1} + c|\hat\lambda_{p-1}] \nonumber \\ 
    &= E[x_p + a(\hat\lambda_{p-1} - \epsilon_p) + c|\hat\lambda_{p-1}] \nonumber \\
    \label{eq:meanlambda}
    &= a\hat\lambda_{p-1} + c.
\end{align}

The other terms require multiple expressions that are more complicated. These expressions include
\begin{align}
    E[\eta_p|\hat\eta_p,\hat\lambda_{p-1}] &= E[\hat\eta_p - \gamma_p|\hat\eta_p] \nonumber \\
\label{eq:meaneta}
    &= \hat\eta_p, \\ \nonumber \\
    E[\lambda_p^2|\hat\eta_p,\hat\lambda_{p-1}] &= E[\lambda_p^2|\hat\lambda_{p-1}] \nonumber \\
    &= E[(x_p + a\lambda_{p-1} + c)^2|\hat\lambda_{p-1}] \nonumber \\ 
    &= E[x_p^2 + 2ax_p\lambda_{p-1} + 2cx_p \nonumber \\
    & \quad + 2ac\lambda_{p-1} + a^2\lambda_{p-1}^2 + c^2|\hat\lambda_{p-1}] \nonumber \\ 
    &= E[x_p^2  + 2ac(\hat\lambda_{p-1} - \epsilon_p) \nonumber \\
    & \quad + a^2(\hat\lambda_{p-1} - \epsilon_p)^2 + c^2|\hat\lambda_{p-1}] \nonumber \\ 
    &= \sigma^2_x + 2acE[\hat\lambda_{p-1} - \epsilon_p|\hat\lambda_{p-1}] \nonumber \\
    & \quad + a^2E[\hat\lambda_{p-1}^2 - 2\epsilon_p\hat\lambda_{p-1} + \epsilon_p^2|\hat\lambda_{p-1}] + c^2 \nonumber \\ 
\label{eq:meanlambdasquared}
    &= \sigma^2_x + 2ac\hat\lambda_{p-1} + a^2\hat\lambda_{p-1}^2 + a^2\sigma^2_{\epsilon} + c^2,
\end{align}
and
\begin{align}
    E[\eta_p^2|\hat\eta_p,\hat\lambda_{p-1}] &= E[(\hat\eta_p - \gamma_p)^2|\hat\eta_p] \nonumber \\
    &= E[\hat\eta_p^2 - 2\hat\eta_p\gamma_p + \gamma_p^2|\hat\eta_p] \nonumber \\
\label{eq:meanetasquared}
    &= \hat\eta_p^2 + \sigma^2_{\gamma}.
\end{align}
By using (\ref{eq:meanlambdasquared}) and (\ref{eq:meanlambda}),
\begin{align}
    \varrm(\lambda_p|\hat\eta_p,\hat\lambda_{p-1}) &= \varrm(\lambda_p|\hat\lambda_{p-1}) \nonumber \\
    &= E[\lambda_p^2|\hat\lambda_{p-1}] - (E[\lambda_p|\hat\lambda_{p-1}])^2 \nonumber \\
    &= \sigma^2_x + 2ac\hat\lambda_{p-1} + a^2\hat\lambda_{p-1}^2 + a^2\sigma^2_{\epsilon} + c^2 \nonumber \\
    & \quad - (a\hat\lambda_{p-1} + c)^2 \nonumber \\
\label{eq:varlambda}
    &= \sigma^2_x + a^2\sigma^2_{\epsilon},
\end{align}
and by using (\ref{eq:meanetasquared}) and (\ref{eq:meaneta}),
\begin{align}
    \varrm(\eta_p|\hat\eta_p,\hat\lambda_{p-1}) &= \varrm(\eta_p|\hat\eta_p) \nonumber \\
    &= E[\eta_p^2|\hat\eta_p] - (E[\eta_p|\hat\eta_p])^2 \nonumber \\
    &= \hat\eta_p^2 + \sigma^2_{\gamma} - \hat\eta_p^2 \nonumber \\
\label{eq:vareta}
    &= \sigma^2_{\gamma}.
\end{align}
Moving forward, by using (\ref{eq:basicymean}) and (\ref{eq:meaneta}) we can compute
\begin{align}
    E[(y_p|\hat\eta_p,\hat\lambda_{p-1})|\lambda_p]
    &= E[E[((y_p|\hat\eta_p,\hat\lambda_{p-1})|\lambda_p)|\eta_p]] \nonumber \\
    &= E[(\lambda_p\eta_p|\hat\eta_p,\hat\lambda_{p-1})|\lambda_p] \nonumber \\
    &= \lambda_pE[\eta_p|\hat\eta_p,\hat\lambda_{p-1}] \nonumber \\
\label{eq:meanygivenlambda}
    &= \lambda_p\hat\eta_p,
\end{align}
and by using (\ref{eq:meanygivenlambda}) and (\ref{eq:meanlambdasquared}) we have
\begin{align}
    E[\lambda_p&y_p|\hat\eta_p,\hat\lambda_{p-1}] = E[E[(\lambda_py_p|\hat\eta_p,\hat\lambda_{p-1})|\lambda_p]] \nonumber \\
    &= E[\lambda_pE[(y_p|\hat\eta_p,\hat\lambda_{p-1})|\lambda_p]] \nonumber \\
    &= E[\lambda_p^2\hat\eta_p|\hat\eta_p,\hat\lambda_{p-1}] \nonumber \\ 
    &= \hat\eta_pE[\lambda_p^2|\hat\eta_p,\hat\lambda_{p-1}] \nonumber \\
\label{eq:meanlambday}    
    &= \hat\eta_p(b^2\sigma^2_x + 2ac\hat\lambda_{p-1} + a^2\hat\lambda_{p-1}^2 + a^2\sigma^2_{\epsilon} + c^2).
\end{align}
Now that we have these terms, we can use (\ref{eq:meanygivenlambda}) and (\ref{eq:meanlambda}) to derive the second term needed for the sequential filter:
\begin{align}
    E[y_p|\hat\eta_p,\hat\lambda_{p-1}] &= E[E[(y_p|\hat\eta_p,\hat\lambda_{p-1})|\lambda_p]] \nonumber \\
    &= E[\lambda_p\hat\eta_p|\hat\eta_p,\hat\lambda_{p-1}] \nonumber \\
    &= \hat\eta_pE[\lambda_p|\hat\eta_p,\hat\lambda_{p-1}] \nonumber \\
\label{eq:meany}
    &= \hat\eta_p(a\hat\lambda_{p-1} + c).
\end{align}

In order to compute the third term, we again need multiple expressions. By using (\ref{eq:basicyvar}), (\ref{eq:meaneta}), and (\ref{eq:meanetasquared}), we have
\begin{align}
    E[\varrm(((y_p|\hat\eta_p,&\hat\lambda_p)|\lambda_p)|\eta_p)] \nonumber \nonumber \\
    &= E[(\lambda_p(\eta_p + \eta_p^2)|\hat\eta_p,\hat\lambda_p)|\lambda_p] \nonumber \\
    &= \lambda_pE[\eta_p + \eta_p^2|\hat\eta_p,\hat\lambda_p] \nonumber \\
\label{eq:meanvarygivenlambdaeta}
    &= \lambda_p(\hat\eta_p + \hat\eta_p^2 + \sigma^2_{\gamma}).
\end{align}
By using (\ref{eq:basicymean}) and (\ref{eq:vareta}),
\begin{align}
    \varrm(E[((y_p|\hat\eta_p,&\hat\lambda_{p-1})|\lambda_p)|\eta_p]) \nonumber \\
    &= \varrm((\lambda_p\eta_p|\hat\eta_p,\hat\lambda_{p-1})|\lambda_p) \nonumber \\ 
    &= \lambda_p^2\varrm(\eta_p|\hat\eta_p,\hat\lambda_{p-1}) \nonumber \\
\label{eq:varmeanygivenlambdaeta}
    &= \lambda_p^2\sigma^2_{\gamma},
\end{align}
and by using the law of total variance with (\ref{eq:meanvarygivenlambdaeta}) and (\ref{eq:varmeanygivenlambdaeta}),
\begin{align}
    \varrm((y_p|\hat\eta_p,&\hat\lambda_{p-1})|\lambda_p) \nonumber \\
    &= E[\varrm(((y_p|\hat\eta_p,\hat\lambda_{p-1})|\lambda_p)|\eta_p)] \nonumber \\
    & \quad + \varrm(E[((y_p|\hat\eta_p,\hat\lambda_{p-1})|\lambda_p)|\eta_p]) \nonumber \\
\label{eq:varygivenlambda}
    &= \lambda_p(\hat\eta_p + \hat\eta_p^2 + \sigma^2_{\gamma}) + \lambda_p^2\sigma^2_{\gamma}.
\end{align}
Continuing on, by using (\ref{eq:varygivenlambda}), (\ref{eq:meanlambda}), and (\ref{eq:meanlambdasquared}),
\begin{align}
    E[&\varrm((y_p|\hat\eta_p,\hat\lambda_{p-1})|\lambda_p)] \nonumber \\
    &= E[\lambda_p(\hat\eta_p + \hat\eta_p^2 + \sigma^2_{\gamma}) + \lambda_p^2\sigma^2_{\gamma}|\hat\eta_p,\hat\lambda_{p-1}] \nonumber \\
    &= (\hat\eta_p + \hat\eta_p^2 + \sigma^2_{\gamma})E[\lambda_p|\hat\eta_p,\hat\lambda_{p-1}] + \sigma^2_{\gamma}E[\lambda_p^2|\hat\eta_p,\hat\lambda_{p-1}] \nonumber \\
\label{eq:meanvarygivenlambda}
    &= (\hat\eta_p + \hat\eta_p^2 + \sigma^2_{\gamma})(a\hat\lambda_{p-1} + c) \nonumber \\
    & \quad + \sigma^2_{\gamma}(\sigma^2_x + 2ac\hat\lambda_{p-1} + a^2\hat\lambda_{p-1}^2 + a^2\sigma^2_{\epsilon} + c^2),
\end{align}
and by using (\ref{eq:meanygivenlambda}) and (\ref{eq:varlambda}),
\begin{align}
    \varrm(E[(y_p|\hat\eta_p,\hat\lambda_p)|\lambda_p]) &= \varrm(\lambda_p\hat\eta_p|\hat\eta_p,\hat\lambda_p) \nonumber \\
    &= \hat\eta_p^2\varrm(\lambda_p|\hat\eta_p,\hat\lambda_p) \nonumber \\
\label{eq:varmeanygivenlambda}
    &= \hat\eta_p^2(\sigma^2_x + a^2\sigma^2_{\epsilon}).
\end{align}
By using (\ref{eq:meanvarygivenlambda}) and
(\ref{eq:varmeanygivenlambda}), we can finally derive our third term for the sequential filter,
\begin{align}
    \varrm(y_p|\hat\eta_p,\hat\lambda_p) &= E[\varrm((y_p|\hat\eta_p,\hat\lambda_{p-1})|\lambda_p)] \nonumber \\
    & \quad + \varrm(E[(y_p|\hat\eta_p,\hat\lambda_p)|\lambda_p]) \nonumber \\
\label{eq:vary}
    &= (\hat\eta_p + \hat\eta_p^2 + \sigma^2_{\gamma})(a\hat\lambda_{p-1} + c) \nonumber \\
    & \quad + \sigma^2_{\gamma}(\sigma^2_x + 2ac\hat\lambda_{p-1} + a^2\hat\lambda_{p-1}^2 + a^2\sigma^2_{\epsilon} + c^2) \nonumber \\
    & \quad + \hat\eta_p^2(\sigma^2_x + a^2\sigma^2_{\epsilon}).
\end{align}

To compute the fourth and final term, by using (\ref{eq:meanlambda}) and (\ref{eq:meany}), we can first compute
\begin{align}
    E[\lambda_p|\hat\eta_p,\hat\lambda_{p-1}]E[y_p|&\hat\eta_p,\hat\lambda_{p-1}] = \hat\eta_p(a\hat\lambda_{p-1} + c)^2 \nonumber \\ 
\label{eq:meanlambdameany}
    &= \hat\eta_p(a^2\hat\lambda_{p-1}^2 + 2ac\hat\lambda_{p-1} + c^2).
\end{align}
Using (\ref{eq:meanlambday}) and (\ref{eq:meanlambdameany}), we can finally get the last term we need:
\begin{align}
    \covrm(\lambda_p,y_p|&\hat\eta_p,\hat\lambda_{p-1}) \nonumber \\
    &= E[\lambda_py_p|\hat\eta_p,\hat\lambda_{p-1}] \nonumber \\
    & \quad - E[\lambda_p|\hat\eta_p,\hat\lambda_{p-1}]E[y_p|\hat\eta_p,\hat\lambda_{p-1}] \nonumber \\
    &= \hat\eta_p(\sigma^2_x + 2ac\hat\lambda_{p-1} + a^2\hat\lambda_{p-1}^2 + a^2\sigma^2_{\epsilon} + c^2) \nonumber \\
    & \quad - \hat\eta_p(a^2\hat\lambda_{p-1}^2 + 2ac\hat\lambda_{p-1} + c^2) \nonumber \\
\label{eq:covlambday}
    &= \hat\eta_p(\sigma^2_{x} + a^2\sigma^2_{\epsilon}).
\end{align}

Now that all of the required terms have been derived, they can all be put together to give us the complete sequential filter. By using (\ref{eq:covlambday}), (\ref{eq:vary}), (\ref{eq:meany}), and (\ref{eq:meanlambda}),
\begin{align}
\label{eq:finalfilter}
    \hat{\lambda}_{p}(y_{p}) =
    &\frac{\hat{\eta}_{p}(b^2\sigma^2_x + a^2\sigma^2_{\epsilon})}{\left[ \splitfrac{\splitfrac{(a\hat{\lambda}_{p-1} + c)(\hat{\eta}_{p} + \hat{\eta}_{p}^2 + \sigma^2_{\gamma})}{ + \hat{\eta}_{p}^2(b^2\sigma^2_x + a^2\sigma^2_{\epsilon})}}{\splitfrac{+ \sigma^2_{\gamma}(b^2\sigma^2_x + a^2\sigma^2_{\epsilon} + a^2\hat{\lambda}_{p-1}^2}{ + 2ac\hat{\lambda}_{p-1} + c^2)}} \right]}
    \nonumber \\
    &\times \: (y_{p} - \hat{\eta}_{p}(a\hat{\lambda}_{p-1} + c)) \: + \: (a\hat{\lambda}_{p-1} + c).
\end{align}

\section*{Acknowledgment}

I would like to thank my thesis advisor Vivek Goyal for advising and teaching me, Sheila Seidel for mentoring me, Minxu Peng for guiding me and laying the groundwork for my project, Akshay Agarwal for educating me about real-world PBM situations, and the entire research group for all their help over the past 1.5 years. I have definitely learned a lot, while also enjoying my time working with them.

\bibliographystyle{IEEEtran}
\bibliography{bibfile}

\end{document}